\documentclass[superscriptaddress,eqsecnum,floats,aps,nofootinbib,showpacs,preprintnumbers]{revtex4}

\usepackage{amsmath,amssymb,amsfonts}
\usepackage{graphicx}

\def\be{\begin{equation}}
\def\ee{\end{equation}}
\def\ba{\begin{eqnarray}}
\def\ea{\end{eqnarray}}
\def\nn{\nonumber}

\newcommand{\abs}[1]{{\left|{#1}\right|}} 

\newcommand{\ftriad}[2]{{}^o\! e^{#1}_{#2}} 
\newcommand{\fcotriad}[2]{{}^o\!\omega_{#1}^{#2}} 
\newcommand{\fq}{{}^o\!q} 
\newcommand{\tE}[2]{{\mbox{$\tilde{E}$}^{#1}}_{#2}} 

\newcommand{\Pl}{\ell_{\rm Pl}} 
\newcommand{\muzero}{\mu^o} 
\newcommand{\mubar}{{\bar \mu}} 

\newcommand{\secref}[1]{Sec.~\ref{#1}}

\newcommand{\eqnref}[1]{(\ref{#1})}
\newcommand{\figref}[1]{Fig.~\ref{#1}}
\newcommand{\tabref}[1]{Table~\ref{#1}}
\newcommand{\appref}[1]{Appendix~\ref{#1}}
\newcommand{\footref}[1]{footnote~\ref{#1}}

\begin{document}


\preprint{IGC-08/7-1}

\title{Phenomenological Loop Quantum Geometry of the Schwarzschild Black Hole}

\author{Dah-Wei Chiou}
\email{chiou@gravity.psu.edu}
\affiliation{
Institute for Gravitation and the Cosmos,
Physics Department,
The Pennsylvania State University, University Park, PA 16802, USA}

\begin{abstract}
The interior of a Schwarzschild black hole is investigated at the level of phenomenological dynamics with the discreteness corrections of loop quantum geometry implemented in two different improved quantization schemes. In one scheme, the classical black hole singularity is resolved by the quantum bounce, which bridges the black hole interior with a white hole interior. In the other scheme, the classical singularity is resolved and the event horizon is also diffused by the quantum bounce. Jumping over the quantum bounce, the black hole gives birth to a baby black hole with a much smaller mass. This lineage continues as each classical black hole brings forth its own descendant in the consecutive classical cycle, giving the whole extended spacetime fractal structure, until the solution eventually descends into deep Planck regime, signaling a breakdown of the semiclassical description. The issues of scaling symmetry and no-hair theorem are also discussed.
\end{abstract}


\maketitle



\section{Introduction}
It has long been suggested that the singularities in general relativity signal a breakdown of the classical theory and should be resolved by the quantum effects of gravity. Loop quantum gravity (LQG) is one of such candidate theories of quantum gravity and its application to cosmological models is known as loop quantum cosmology (LQC) (see \cite{Bojowald:2006da} for a review). The comprehensive formulation for LQC has been constructed in detail in the spatially flat and isotropic model with a free massless scalar field \cite{Ashtekar:2006uz,Ashtekar:2006rx,Ashtekar:2006wn}, showing that the quantum evolution is deterministic across the deep Planck regime and the cosmological singularity is replaced by a quantum bounce for the states which are semiclassical at late times. This construction was extended to $k=\pm1$ Friedmann-Robertson-Walker models to include intrinsic curvature \cite{Ashtekar:2006es,Vandersloot:2006ws} as well as Bianchi I models to include anisotropy \cite{Chiou:2006qq,Chiou:2007dn,Chiou:2007sp,Chiou:2007mg}, affirming the resolution of cosmological singularities and the occurrence of quantum bounces, either in the fundamental quantum theory of LQC or at the level of phenomenological dynamics.

To further extend this formulation and enlarge its domain of validity, the next step is to investigate loop quantum geometry of the black hole and to see whether the black hole singularity is also resolved. The simplest step is to consider the interior of a Schwarzschild black hole, in which the temporal and radial coordinates flip roles and thus the metric components are homogeneous with the Kantowski-Sachs symmetry. Thanks to homogeneity, the loop quantization of the Schwarzschild interior can be formulated as a minisuperspace model in a similar fashion to LQC. This has been developed in \cite{Modesto:2004wm,Ashtekar:2005qt,Modesto:2005zm} and its phenomenological dynamics studied in \cite{Modesto:2006mx} shows that the black hole interior is extended to a white hole interior through the bounce, which resolves the singularity.

The analysis in \cite{Modesto:2006mx} is based on the original quantization strategy (referred to as the ``$\mu_o$-scheme'' in this paper) used in \cite{Ashtekar:2005qt}, which, as a direct transcription of the original LQC construction in \cite{Ashtekar:2006uz}, introduces a fixed parameter to impose fundamental discreteness of quantum geometry. However, it has been argued that the $\mu_o$-scheme quantization in LQC leads to a wrong semiclassical limit in some regimes and should be improved by replacing the discreteness parameters with adaptive variables which depend on the scale factors \cite{Ashtekar:2006wn}. Two improved strategies (called ``$\mubar$-scheme'' and ``$\mubar'$-scheme'' in this paper) for loop quantization of the Schwarzschild interior were investigated in \cite{Bohmer:2007wi} at the level of phenomenological dynamics.

However, the results of \cite{Bohmer:2007wi} are not easily compared with the bouncing scenario of LQC as some details are still missing. For instance, the exact condition for the occurrence of the bounce has yet to be pinpointed. To have a better understanding of the extended Schwarzschild solution, more effort is needed to investigate the quantum corrections on the horizon and the evolution of the parameters (e.g., mass of the black hole) that characterize different classical phases across the quantum bounce.

In order to bridge the gap between the loop quantum dynamics of cosmological models and that of Schwarzschild black holes, a cosmological model of Kantowski-Sachs spacetime with a massless scalar field has been studied in \cite{Chiou:2008eg} at the level of phenomenological dynamics. The study of \cite{Chiou:2008eg} not only sets a new cosmological model of LQC with inclusion of both intrinsic curvature and anisotropy but also facilitates a methodology to study the details of loop quantum geometry of the Schwarzschild interior. By exploiting the methods introduced in \cite{Chiou:2008eg}, we are able to articulate the geometrical interpretation of the extended Schwarzschild solution and refine some observations obtained in \cite{Bohmer:2007wi}.

Based on the same semiclassical approach of \cite{Chiou:2008eg} to incorporate loop quantum corrections, the phenomenological dynamics of the Schwarzschild interior is investigated in this paper with two improved quantization strategies ($\mubar$- and $\mubar'$-schemes). In the $\mubar$-scheme, the classical singularity is resolved and replaced by the quantum
bounce, which bridges the black hole interior with the interior of a white hole. On the other hand, in the $\mubar'$-scheme, the classical black hole singularity is resolved
and the event horizon is diffused by the quantum bounce, across which, the classical black hole gives birth to a baby black hole with a decreased mass in the consecutive classical cycle. This lineage continues, giving the extended spacetime ``fractal'' structure, until eventually the triad variable $p_b$ grows exponentially while the other triad variable $p_c$ descends into a deep Planck regime, signaling a breakdown of the semiclassical description.

As in the Hamiltonian framework for homogeneous models, we have to restrict the spatial integration to a finite sized shell $\mathcal{I}\times S^2$ to make the Hamiltonian finite. This prescription raises the question whether the resulting dynamics is independent of the choice of $\mathcal{I}$. It can be shown that the phenomenological dynamics in the $\mubar'$-scheme is completely independent of the choice of $\mathcal{I}$ as is the classical dynamics, while the phenomenological dynamics in the $\mubar$-scheme reacts to $\mathcal{I}$ and thus, in the language of the ``no-hair'' theorem, one extra parameter (mass of the conjoined white hole) is required to completely characterize the extended Schwarzschild solution.

In addition to the issues related to the dependence on $\mathcal{I}$, the phenomenological dynamics also reveals interesting scaling symmetry, which is suggestive that the fundamental scale (area gap) imposed for the spatial geometry may give rise to a fundamental scale in temporal measurement.

This paper follows the steps in \cite{Chiou:2008eg} as closely as possible and uses the same notations thereof.\footnote{Assiduous readers are encouraged to look at \cite{Chiou:2007mg} and \cite{Chiou:2008eg} to see the close parallels.} In \secref{sec:classical dynamics}, the Ashtekar variables with the Kantowski-Sachs symmetry are introduced and the classical geometry of the Schwarzschild interior is solved in Hamiltonian formalism. The phenomenological dynamics with discreteness corrections of loop quantum geometry is constructed and solved in \secref{sec:phenomenological dynamics} for the $\mubar$- and $\mubar'$-schemes, respectively. The scaling symmetry and related issues are discussed in \secref{sec:scaling}. Finally, the results are summarized and discussed in \secref{sec:summary}. For comparison, the phenomenological dynamics in the $\mu_o$-scheme is also included in \appref{sec:muzero dynamics}.

\section{Classical dynamics}\label{sec:classical dynamics}
In this section, we first briefly describe the Ashtekar variables for the geometry invariant under the Kantowski-Sachs symmetry \cite{Ashtekar:2005qt}. In the Hamiltonian framework, we then solve the classical solution in terms of Ashtekar variables for the interior of a Schwarzschild black hole.

\subsection{Ashtekar variables with the Kantowski-Sachs symmetry}\label{sec:Ashtekar variables}
The metric of homogeneous spacetime with the Kantowski-Sachs symmetry group $\mathbb{R}\times SO(3)$ is given by the line element:
\ba\label{eqn:metric}
ds^2&=&-d\tau^2+g_{xx}(\tau)dx^2+g_{\Omega\Omega}(\tau)d\Omega^2\nn\\
&=&-N(t)^2dt^2+g_{xx}(t)dx^2+g_{\theta\theta}(t)d\theta^2+g_{\phi\phi}(t)d\phi^2,
\ea
where $\tau$ is the proper time, $N(t)$ is the lapse function associated with the arbitrary coordinate time $t$ via $N(t)dt=d\tau$ and $d\Omega^2$ represents the unit 2-sphere given in polar coordinates as
\be
d\Omega^2=d\theta^2+\sin^2\theta\,d\phi^2.
\ee
The topology of the homogeneous spatial slices is $\Sigma=\mathbb{R}\times S^2$, which is coordinatized by $x\in\mathbb{R}$, $\theta\in[0,\pi]$ and $\phi\in[0,2\pi]$.

As in any homogeneous cosmological models, on the homogeneous spacelike slice $\Sigma$, we can choose a fiducial triad field of vectors $\ftriad{a}{i}$ and a fiducial cotriad field of covectors $\fcotriad{a}{i}$ that are left-invariant by the action of the Killing fields of $\Sigma$. (Note $\ftriad{a}{i}\fcotriad{b}{i}=\delta^a_b$.) The \emph{fiducial} 3-metric of $\Sigma$ is given by the cotriad $\fcotriad{a}{i}$:
\be
\fq_{ab}=\fcotriad{a}{i}\,\fcotriad{b}{j}\,\delta_{ij}.
\ee
In the \emph{comoving coordinates} $(x,\theta,\phi)$, we can choose $\fq_{ab}$ to have
\be
\fq_{ab}dx^adx^b=dx^2+d\theta^2+\sin^2\theta\, d\phi^2,
\ee
which gives $\fq:=\det \fq_{ab}=\sin^2\theta$.

In connection dynamics, the canonical pair consists of the Ashtekar variables:
the densitized triads $\tE{a}{i}(\vec{x})$ and connections  ${A_a}^i(\vec{x})$, which satisfy the canonical relation:
\be\label{eqn:Poisson of A nd E}
\{{A_a}^i(\vec{x}),\tE{b}{j}(\vec{x}')\}
=8\pi G\gamma\,\delta^i_j\,\delta_a^b\,\delta^3(\vec{x}-\vec{x}'),
\ee
where $\gamma$ is the Barbero-Immirzi parameter. In the case that connections and triads admit the Kantowski-Sachs symmetry $\mathbb{R}\times SO(3)$, ${A_a}^i$ and $\tE{a}{i}$ after gauge fixing of the Gauss constraint are of the form \cite{Ashtekar:2005qt}:
\ba
\label{eqn:symmetric A}
A={A_a}^i\tau_idx^a&=&\tilde{c}\tau_3dx+\tilde{b}\tau_2d\theta
-\tilde{b}\tau_1\sin\theta d\phi+\tau_3\cos\theta d\phi,\\
\label{eqn:symmetric E}
\tilde{E}=\tE{a}{i}\tau_i\partial_a&=&\tilde{p}_c\tau_3\sin\theta\,\partial_x
+\tilde{p}_b\tau_2\sin\theta\,\partial_\theta-\tilde{p}_b\tau_1\,\partial_\phi,
\ea
where $\tilde{b}$, $\tilde{c}$, $\tilde{p}_b$, $\tilde{p}_c$ are functions of time only and $\tau_i=-i\sigma_i/2$ are $SU(2)$ generators satisfying $[\tau_i,\tau_i]={\epsilon_{ij}}^k\tau_k$ (with $\sigma_i$ being the Pauli matrices.)

The symplectic structure on the symmetry-reduced phase space is given by the complete symplectic structure [as in \eqnref{eqn:Poisson of A nd E}] integrated over the finite sized shell $\mathcal{I}\times S^2$:
\be\label{eqn:Omega tilde}
\tilde{{\bf \Omega}}=\frac{1}{8\pi G\gamma}\int_{\mathcal{I}\times S^2}d^3x
\;d{A_a}^i(\vec{x})\wedge d\tE{a}{i}(\vec{x})
=\frac{L}{2G\gamma}
\left(d\tilde{c}\wedge d\tilde{p}_c+2d\tilde{b}\wedge d\tilde{p}_b\right),
\ee
where the integration is over $\theta\in[0,\pi]$, $\phi\in[0,2\pi]$ and restricted to $x\in\mathcal{I}:=[0,L]$; the finite interval $\mathcal{I}$ is prescribed to circumvent the problem due to homogeneity that the spatial integration over the whole spatial slice $\mathbb{R}\times S^2$ diverges. (We will see that this prescription does not change the classical dynamics but might have effects on the quantum corrections.)
The reduced symplectic form leads to the canonical relations for the reduced canonical variables:
\be
\{\tilde{b},\tilde{p}_b\}=G\gamma L^{-1},
\qquad
\{\tilde{c},\tilde{p}_c\}=2G\gamma L^{-1}
\ee
and $\{\tilde{b},\tilde{c}\}=\{\tilde{p}_b,\tilde{p}_c\}=0$.
It is convenient to introduce the rescaled variables:
\be
b:=\tilde{b},\qquad
c:=L\tilde{c},\qquad
p_b:=L\tilde{p}_b,\qquad
p_c:=\tilde{p}_c,
\ee
which satisfy the canonical relations:
\be
\{b,p_b\}=G\gamma,
\qquad
\{c,p_c\}=2G\gamma.
\ee

The relation between the densitized triad and the 3-metric is given by
\be\label{eqn:q and E}
qq^{ab}=\delta^{ij}\tE{a}{i}\tE{b}{j},
\ee
which leads to
\be
g_{\Omega\Omega}=g_{\theta\theta}=g_{\phi\phi}\sin^2\theta=p_c,\qquad
g_{xx}=\frac{p_b^2}{L^2p_c}.
\ee
Let $S_{x\phi}$, $S_{x\theta}$ and $S_{\theta\phi}$ be the three surfaces of interest, respectively, bounded by the interval $\mathcal{I}$ and the equator, $\mathcal{I}$ and a great circle along a longitude, and the equator and a longitude (so that $S_{\theta\phi}$ forms a quarter of the sphere $S^2$). It follows that the physical areas of $S_{x\phi}$, $S_{x\theta}$ and $S_{\theta\phi}$ are given by
\be\label{eqn:p and A}
{\bf A}_{x\phi}={\bf A}_{x\theta}=2\pi L\sqrt{g_{xx}g_{\Omega\Omega}}=2\pi p_b,
\qquad
{\bf A}_{\theta\phi}=\pi g_{\Omega\Omega}=\pi p_c,
\ee
and the physical volume of $\mathcal{I}\times S^2$ is
\be\label{eqn:V}
{\bf V}=4\pi L \sqrt{g_{xx}}\,g_{\Omega\Omega}=4\pi p_b \sqrt{p_c}\,.
\ee
This gives the physical meanings of the triad variables $p_b$ and $p_c$.\footnote{More precisely, in \eqnref{eqn:p and A} $p_b$ and $p_c$ should be $\abs{p_b}$ and $\abs{p_c}$ \cite{Ashtekar:2005qt}. With the gauge fixing $p_b>0$, the opposite sign of $p_c$ corresponds to the inverse spatial orientation, which we do not need to consider on this paper.}

\subsection{Classical solution}\label{sec:classical solution}


The vacuum solution of Kantowski-Sachs spacetime is identified with the interior of a Schwarzschild black hole (as will be shown in \secref{sec:black hole interior}). The Hamiltonian constraint of the Schwarzschild interior is given in terms of Ashtekar variables as
\be\label{eqn:cl Hamiltonian}
H=-\frac{N}{2G\gamma^2}\left[
2bc\sqrt{p_c}+(b^2+\gamma^2)\frac{p_b}{\sqrt{p_c}}
\right].
\ee
This can be derived from the Hamiltonian constraint of the full theory of LQG. (See \cite{Ashtekar:2005qt} or Appendix B of \cite{Chiou:2008eg}.)

To solve the classical solution, we can simplify the Hamiltonian by choosing the lapse function $N=p_b\sqrt{p_c}\equiv{\bf V}/4\pi$ and thus introducing the conformal time variable $dt'=(p_b\sqrt{p_c})^{-1}d\tau$.
The rescaled Hamiltonian is given by
\be\label{eqn:cl rescaled Hamiltonian}
H'=-\frac{1}{2G\gamma^2}\left[2bcp_bp_c+(b^2+\gamma^2)p_b^2\right].
\ee

The equations of motion are governed by the Hamilton's equations:
\ba
\label{eqn:cl eom 3}
\frac{dc}{dt'}&=&\{c,H'\}=2G\gamma\,\frac{\partial\, H'}{\partial p_c}=-2\gamma^{-1}c b p_b,\\
\label{eqn:cl eom 4}
\frac{dp_c}{dt'}&=&\{p_c,H'\}=-2G\gamma\,\frac{\partial\, H'}{\partial c}=2\gamma^{-1}p_c b p_b,\\
\label{eqn:cl eom 5}
\frac{db}{dt'}&=&\{b,H'\}=G\gamma\,\frac{\partial\, H'}{\partial p_b}=-b\gamma^{-1}\left(bp_b+cp_c\right)-\gamma p_b,\\
\label{eqn:cl eom 6}
\frac{dp_b}{dt'}&=&\{p_b,H'\}=-G\gamma\,\frac{\partial\, H'}{\partial b}= \gamma^{-1}p_b\left(bp_b+cp_c\right),
\ea
as well as the constraint that the Hamiltonian must vanish:
\be\label{eqn:cl eom 7}
H'=0\quad\Rightarrow\quad
2bcp_bp_c+\left(b^2+\gamma^2\right)p_b^2=0.
\ee
Notice that substituting \eqnref{eqn:p and A} into \eqnref{eqn:cl eom 4} and \eqnref{eqn:cl eom 6} gives us
\ba
\label{eqn:cl b}
b&=&\frac{\gamma}{2\sqrt{p_c}}\frac{dp_c}{d\tau}
=\gamma\frac{d}{d\tau}\sqrt{g_{\Omega\Omega}}\,,\\
\label{eqn:cl c}
c&=&\frac{\gamma}{p_c^{1/2}}\frac{dp_b}{d\tau}-\frac{\gamma p_b}{2p_c^{3/2}}\frac{dp_c}{d\tau}
=\gamma\frac{d}{d\tau}\left(\frac{p_b}{\sqrt{p_c}}\right)
=\gamma\frac{d}{d\tau}\left(L\sqrt{g_{xx}}\right),
\ea
which tells that, \emph{classically}, the connection variable $b$ is the time change rate of the square root of the physical area of $S^2$ [up to constant $(4\pi)^{-1}\gamma$] and $c$ is the time change rate of the physical length of $\mathcal{I}$ (up to constant $\gamma$).

To solve the equations of motion, first note that combining \eqnref{eqn:cl eom 3} and \eqnref{eqn:cl eom 4} gives
\ba\label{eqn:Kc}
\frac{d}{dt'}(p_c c)=0\quad\Rightarrow\quad
p_c c=\gamma K_c\ \text{is constant},
\ea
and on the other hand, \eqnref{eqn:cl eom 5} and \eqnref{eqn:cl eom 6} yield
\ba\label{eqn:Kb}
\frac{d}{dt'}(K_b)=-p_b^2\qquad
\text{with}\ p_bb=:\gamma K_b(t').
\ea
The Hamiltonian constraint \eqnref{eqn:cl eom 7} then reads as
\be\label{eqn:Kb Kc and pb}
2K_bK_c+K_b^2+p_b^2=0.
\ee
By \eqnref{eqn:Kb} and \eqnref{eqn:Kb Kc and pb}, we have
\be\label{eqn:cl diff eq for Kb}
\frac{dK_b}{dt'}=2K_bK_c+K_b^2,
\ee
the solution of which is given by
\be\label{eqn:cl sol of Kb}
K_b(t')=-2K_c\,\frac{\alpha e^{2K_c(t'-t'_0)}}{1+\alpha e^{2K_c(t'-t'_0)}}
\ee
with $\alpha$ being a dimensionless constant specified by the initial state:
\be\label{eqn:const alpha}
\alpha:=-\frac{K_b(t'_0)}{K_b(t'_0)+2K_c}.
\ee
In terms of $K_b(t')$ and the constant $K_c$, \eqnref{eqn:cl eom 4} and \eqnref{eqn:cl eom 6} now read as
\ba
\label{eqn:cl diff eq for pc}
\frac{1}{p_c}\frac{dp_c}{dt'}
&\equiv& \frac{{\bf V}}{4\pi p_c}\frac{dp_c}{d\tau}= 2K_b(t'),\\
\label{eqn:cl diff eq for pb}
\frac{1}{p_b}\frac{dp_b}{dt'}
&\equiv& \frac{{\bf V}}{4\pi p_b}\frac{dp_b}{d\tau}=K_b(t')+K_c,
\ea
the solutions to which are given by
\be\label{eqn:cl sol of pc}
p_c(t')=g_{\Omega\Omega}(t')=p_c(t'_0)
\left(\frac{\alpha+1}{\alpha e^{2K_c(t'-t'_0)}+1}\right)^2
\ee
and
\be\label{eqn:cl sol of pb}
p_b(t')=p_b(t'_0)(\alpha+1)\,
\frac{e^{K_c(t'-t'_0)}}{\alpha e^{2K_c(t'-t'_0)}+1}.
\ee
Consequently, we have
\be\label{eqn:cl sol of gxx}
g_{xx}(t')=\frac{p_b^2}{L^2p_c}=\frac{p_b(t'_0)^2}{L^2p_c(t'_0)}
\ e^{2K_c(t'-t'_0)}
\ee
and
\be
{\bf V}(t')=4\pi p_b\sqrt{p_c}=4\pi p_b(t'_0)\sqrt{p_c(t'_0)}\
(\alpha+1)^2\,\frac{e^{K_c(t'-t'_0)}}{\left(\alpha e^{2K_c(t'-t'_0)}+1\right)^2}.
\ee
It should be noted that, by \eqnref{eqn:cl diff eq for pc} and \eqnref{eqn:cl diff eq for pb}, the time reversal $t'\longrightarrow-t'$ corresponds to the sign flipping of $K_c\longrightarrow-K_c$ and $K_b(t')\longrightarrow-K_b(-t')$ simultaneously. For convenience, we fix the convention $K_c>0$ for black holes and $K_c<0$ for white holes. For a black hole, the Hamiltonian constraint \eqnref{eqn:Kb Kc and pb} then yields $K_b(t')<0$ and consequently $\alpha>0$.

With the help of \eqnref{eqn:Kb Kc and pb} and \eqnref{eqn:const alpha}, it can be shown from \eqnref{eqn:cl sol of pb} that $p_b$ reaches the maximal value
\be\label{eqn:pb max}
p_{b,\,\max}=p_b(t'_0)\,\frac{\alpha+1}{2\sqrt{\alpha}}=K_c
\ee
at the epoch $t'=t'_{\rm max}$ satisfying $\alpha \exp(2K_c(t'_{\max}-t'_0))=1$. That is, the constant $2\pi K_c$ can be interpreted as the maximal value of the area ${\bf A}_{x\theta}={\bf A}_{x\phi}\equiv2\pi p_b$.

The solutions of $K_b(t')$, $p_b(t')$ and $p_c(t)$ all approach to constants asymptotically as $t'\rightarrow\pm\infty$. The solutions at the epochs of particular interest are listed as follows:
\be\label{eqn:Kb asymp}
K_b(t')=
\left\{
\begin{array}{lcl}
-2K_c & & \quad\text{as }t'\rightarrow\infty,\\
\,0 & & \quad\text{as }t'\rightarrow-\infty,\\
-K_c & & \quad\text{as }t'=t'_{\max},
\end{array}
\right.
\ee
\be\label{eqn:pb asymp}
p_b(t')=
\left\{
\begin{array}{lcl}
0 & & \quad\text{as }t'\rightarrow\pm\infty,\\
p_{b,\,\max}=K_c & & \quad\text{as }t'=t'_{\max},
\end{array}
\right.
\ee
\be\label{eqn:pc asymp}
p_c(t')=
\left\{
\begin{array}{lcl}
\,0 & & \quad\text{as }t'\rightarrow\infty,\\
p_c(t'_0)(\alpha+1)^2\equiv 4G^2M^2 & & \quad\text{as }t'\rightarrow-\infty,\\
p_c(t'_{\max})=p_c(t'_0)(\alpha+1)^2/4\equiv G^2M^2 & & \quad\text{as }t'=t'_{\max}.
\end{array}
\right.
\ee
Notice that the constant
\ba
\label{eqn:G2M2}
p_c(t'_{\max})&=&G^2M^2\equiv p_c(t'_0)\,\frac{(\alpha+1)^2}{4}
=p_c(t'_0)\left(\frac{K_c}{K_b(t'_0)+2K_c}\right)^2\\
\label{eqn:G2M2'}
&=&p_c(t'_0)\left(\frac{K_b(t'_0)K_c}{p_b(t'_0)^2}\right)^2
\ea
is independent of $t'_0$ [this can be shown by taking the derivative of the right hand side of \eqnref{eqn:G2M2} with respect to $t'_0$ with the help of \eqnref{eqn:cl diff eq for Kb} and \eqnref{eqn:cl diff eq for pc}]. In \secref{sec:black hole interior}, we will identify the constant $p_c(t'_0)(\alpha+1)^2$ as $4G^2M^2$ with $M$ being the mass of the Schwarzschild black hole (i.e., the area of the horizon is given by $16\pi G^2M^2$).

\begin{figure}
\begin{picture}(500,270)(0,0)

\put(-63,-472)
{
\scalebox{1}{\includegraphics{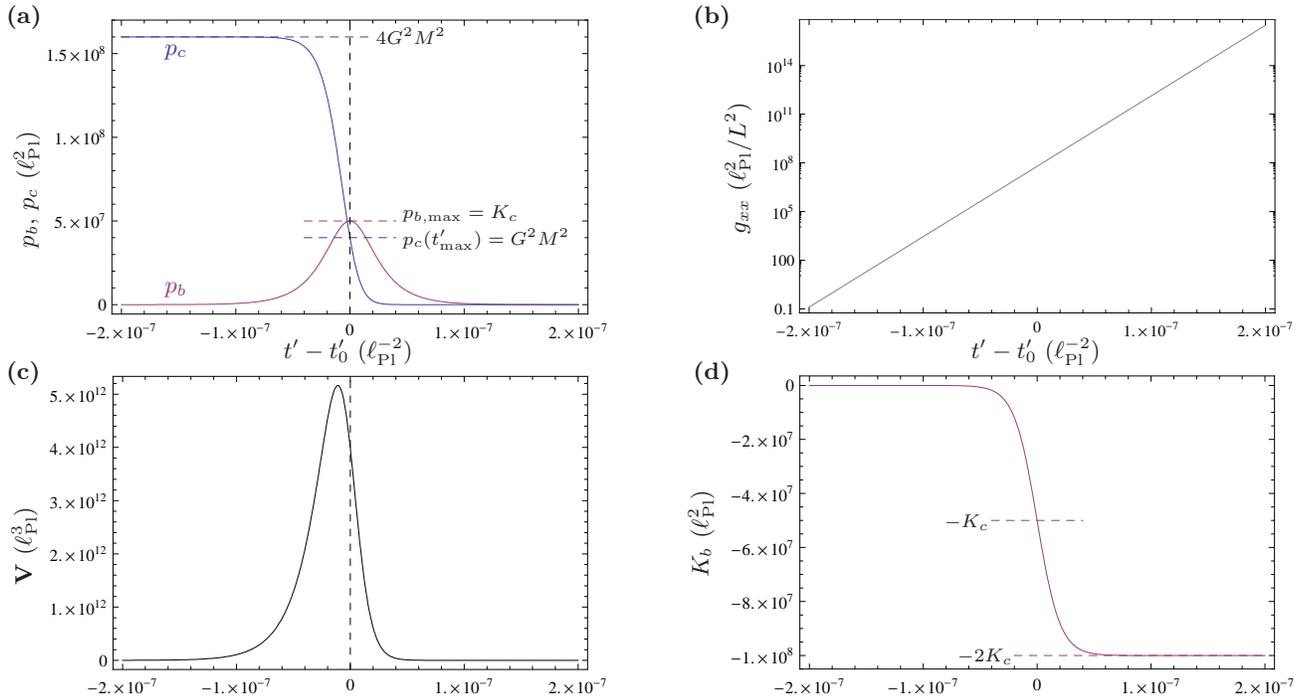}}
}

\end{picture}
\caption{\textbf{Classical solution.} The initial condition is given at $t'_0=t'_{\max}$ (and thus $\alpha=1$) with $p_b(t'_0)=p_{b,\max}=K_c=5.\times10^7\Pl^2$ and $p_c(t'_0)=p_c(t'_{\max})=G^2M^2=4.\times10^7\Pl^2$ ($\Pl:=\sqrt{G\hbar}$ is the Planck length). \textbf{(a)} $p_b(t')$ and $p_c(t')=g_{\Omega\Omega}(t')$. \textbf{(b)} $g_{xx}(t')$. \textbf{(c)} ${\bf V}(t')$. \textbf{(d)} $K_b(t')$.}\label{fig:classical solution}
\end{figure}

Additionally, the asymptotic behavior of $g_{xx}$ is given by
\be
g_{xx}(t')=
\left\{
\begin{array}{ccl}
\infty & & \quad\text{as }t'\rightarrow\infty,\\
0 & & \quad\text{as }t'\rightarrow-\infty,
\end{array}
\right.
\ee
and that of ${\bf V}$ is
\be
{\bf V}(t')=0\quad\text{as }t'\rightarrow\pm\infty.
\ee
The behaviors of the classical solution are depicted in \figref{fig:classical solution}.
In \secref{sec:black hole interior}, we show that the epoch $t'=-\infty$ corresponds to the event horizon of the Schwarzschild black hole and $t'=\infty$ corresponds to the black hole singularity.

Notice that, by \eqnref{eqn:p and A}, \eqnref{eqn:V}, \eqnref{eqn:cl b} and \eqnref{eqn:cl c}, ${\bf V}$, $p_b$ and $c$ depend on the choice of the interval $\mathcal{I}$ and scale as ${\bf V},\,p_b,\,c\propto L$, while $p_c$ and $b$ are independent of $\mathcal{I}$. As a result, the constant of motion $K_c$ as well as the function $K_b(t')$ both scale as $\propto L$. The ratios $K_c/{\bf V}$ and $K_b/{\bf V}$ are nevertheless independent of $\mathcal{I}$; hence \eqnref{eqn:cl diff eq for pc} and \eqnref{eqn:cl diff eq for pb} tell that the differential equations for $p_c^{-1}dp_c/d\tau$ and $p_b^{-1}dp_b/d\tau$ in terms of the proper time $\tau$ are both independent of $\mathcal{I}$. Therefore, the classical dynamics is \emph{completely independent} of the finite interval $\mathcal{I}$ we choose to make sense of the Hamiltonian formalism. (However, the independence of the choice of $\mathcal{I}$ is not necessarily retained when quantum corrections are taken into account.) Furthermore, the black hole mass $M$ is also independent of $\mathcal{I}$ as can be seen on the right-hand side of \eqnref{eqn:G2M2}.

There are 4 degrees of freedom in the phase space of $p_b$, $p_c$, $b$ and $c$. Imposing the Hamiltonian constraint and taking into account the irrelevant choices of the finite interval $\mathcal{I}$ and starting time $t'_0$, we end up with only 4-3=1 genuine degree of freedom. This affirms the ``no-hair theorem'', which states that, in the case of the vacuum solution without angular momentum, stationary, asymptotically flat black holes are uniquely characterized by one parameter of mass. At the level of phenomenological dynamics with loop quantum corrections, the no-hair theorem remains unchanged for the $\mubar'$-scheme but requires one extra parameter (due to the dependence on $\mathcal{I}$) in the $\mubar$-scheme, as will be studied in \secref{sec:mubar dynamics}.

\subsection{Interior of the Schwarzschild black hole}\label{sec:black hole interior}
The standard expression of the Schwarzschild metric in terms of spherical coordinates is given by
\be\label{eqn:Schwarzschild}
ds^2=-\left(1-\frac{2GM}{r}\right)dt^2
+\left(1-\frac{2GM}{r}\right)^{-1}dr^2+r^2d\Omega^2,
\ee
which is asymptotically flat (i.e., as $r\rightarrow\infty$, the metric components approach those of Minkowski spacetime in spherical coordinates.)
Inside the horizon, the temporal and radial coordinates flip roles. To reflect this, we rename the coordinates $(r,t)$ as $(t,\mathcal{L}^{-1} x)$ with an arbitrary scaling factor $\mathcal{L}$. The metric of the Schwarzschild interior now reads as the form of \eqnref{eqn:metric}:
\ba\label{eqn:Schwarzschild interior}
ds^2&=&-N(t)^2dt^2+g_{xx}(t)dx^2+g_{\Omega\Omega}(t)d\Omega^2\nn\\
&=&-\left(\frac{2GM}{t}-1\right)^{-1}dt^2
+\mathcal{L}^{-2}\left(\frac{2GM}{t}-1\right)dx^2+t^2d\Omega^2,
\ea
where $t\in[0,2GM]$, $x\in \mathbb{R}$ and $M$ is the mass of the black hole. The black hole singularity corresponds to $t=0$ and the event horizon corresponds to $t=2GM$. Note that different values of $\mathcal{L}$ correspond to different scalings of $x$ and thus they all give equivalent metric.\footnote{As far as the interior is concerned, $\mathcal{L}$ remains arbitrary. However, if the exterior is also taken into account, there is a canonical convention to fix $\mathcal{L}=1$ such that $t=\mathcal{L}^{-1}x$ in \eqnref{eqn:Schwarzschild} coincides with the proper time in the asymptotically flat regime.}

To show that the solution we get in \secref{sec:classical solution} is the Schwarzschild interior, we first identify $g_{\Omega\Omega}=t^2$. The solution in \eqnref{eqn:cl sol of pc} then yields
\be
e^{2K_c(t'-t'_0)}=\alpha^{-1}
\left(-1+(\alpha+1)\frac{\sqrt{p_c(t'_0)}}{t}\right)
\ee
and
\be
-p_c(t'_0)\,\frac{4K_c\alpha(\alpha+1)^2}{\left(\alpha e^{2K_c(t'-t'_0)}+1\right)^3}\
e^{2K_c(t'-t'_0)}dt'=2tdt.
\ee
Consequently, \eqnref{eqn:cl sol of gxx} reads as
\be
g_{xx}=\frac{4K_c^2}{L^2p_c(t'_0)(\alpha+1)^2}
\left(\frac{(\alpha+1)\sqrt{p_c(t'_0)}}{t}-1\right)
\ee
and the solutions of \eqnref{eqn:cl sol of pc} and \eqnref{eqn:cl sol of pb} give
\ba
d\tau^2&=&N(t')^2d{t'}^2
=\frac{p_b(t'_0)^2}{4K_c^2\alpha\,p_c(t'_0)}\left(
\frac{(\alpha+1)^2p_c(t'_0)}{\frac{(\alpha+1)\sqrt{p_c(t'_0)}}{t}-1}\right)dt^2\nn\\
&=&\left(\frac{(\alpha+1)\sqrt{p_c(t'_0)}}{t}-1\right)^{-1}dt^2
=:N(t)^2dt^2,
\ea
where \eqnref{eqn:pb max} has been used.

If we identify the constant of motion $p_c(t'_{\max})$ in \eqnref{eqn:G2M2} as $G^2M^2$, we then have
\be
N^2(t)=\left(\frac{2GM}{t}-1\right)^{-1}
\ee
and
\be
g_{xx}(t)=\left(\frac{K_c}{LGM}\right)^2\left(\frac{2GM}{t}-1\right),
\ee
which are identical to those in \eqnref{eqn:Schwarzschild interior} with $\mathcal{L}=L(GM/K_c)$.

\section{Phenomenological dynamics with loop quantum corrections}\label{sec:phenomenological dynamics}
In the fundamental loop quantum theory with Kantowski-Sachs symmetry \cite{Ashtekar:2005qt}, the Hamiltonian constraint incorporates two main
sources of corrections: First, the connection variables $b$ and $c$ do not exist and should be replaced by holonomies; second, the discreteness of quantum geometry also modifies the cotriad component $\omega_c:= p_b/\sqrt{p_c}=L\sqrt{g_{xx}}$ such that the eigenvalues of $\hat{\omega}_c$ are finite and significantly different from the classical value near the singularity, at which $p_b/\sqrt{p_c}$ diverges. In the semiclassical description, it is realized that the modification on the cotriad $\omega_c$ is less important. At the level of phenomenological analysis, following the procedures adopted for the isotropic cosmology \cite{Singh:2005xg} and the Bianchi I model \cite{Chiou:2007mg}, we will ignore the correction on $\omega_c$ by simply keeping the classical function $\omega_c=p_b/\sqrt{p_c}$ and take the prescription to replace $b$, $c$ with
\be\label{eqn:sin prescription}
b\longrightarrow\frac{\sin(\mubar_bb)}{\mubar_b},
\qquad
c\longrightarrow\frac{\sin(\mubar_cc)}{\mubar_c},
\ee
introducing the variables $\mubar_b$ and $\mubar_c$ to impose the fundamental discreteness of loop quantum geometry.\footnote{This prescription is sometimes referred to as ``polymerization'' or ``holonomization'' in the literature.}
The heuristic argument for this prescription can be found in Appendix B of \cite{Chiou:2008eg} from the perspective of the full (unreduced) theory of LQG.

With the prescription of \eqnref{eqn:sin prescription} adopted and the cotriad component $\omega_c$ unchanged, by choosing $N=\gamma p_b\sqrt{p_c}$ and $dt'=(\gamma p_b\sqrt{p_c})^{-1}d\tau$, the (rescaled) classical Hamiltonian \eqnref{eqn:cl rescaled Hamiltonian} is modified to serve as the effective Hamiltonian for the semiclassical theory:
\ba\label{eqn:qm Hamiltonian}
H'_\mubar&=&
-\frac{1}{2G\gamma}
\left\{
2\frac{\sin(\mubar_bb)}{\mubar_b}\frac{\sin(\mubar_cc)}{\mubar_c}\,p_bp_c
+\left(\frac{\sin(\mubar_bb)}{\mubar_b}\right)^2 p_b^2
+\gamma^2p_b^2
\right\}.
\ea
The phenomenological dynamics is then solved as if the dynamics was classical but governed by the new effective Hamiltonian.
This treatment is however only heuristic and its validity is still questionable; a more rigorous understanding of the fundamental quantum dynamics would require more sophisticated refinement. Nevertheless, the fact that the phenomenological theory could provide an accurate approximation (for the case where the backreaction is negligible) has been evidenced in the isotropic cosmology \cite{Ashtekar:2006wn,Singh:2005xg,Bojowald:2006gr,Taveras:2008ke} and also affirmed in the Bianchi I model \cite{in progress}.

As for imposing the fundamental discreteness of LQG on the formulation of homogeneous spacetime, the original construction ($\mu_o$-scheme) is to take $\mubar_b$ and $\mubar_c$ as constants (referred to as $\muzero_b, \muzero_c$ in \appref{sec:muzero dynamics} and $\delta$ in \cite{Ashtekar:2005qt,Modesto:2006mx,Bohmer:2007wi}). However, it has been shown in both isotropic and Bianchi I models that the $\mu_o$-scheme can lead to the wrong semiclassical limit\footnote{For the Schwarzschild interior, due to the absence of matter content, it is not obvious whether the $\mu_o$-scheme gives rise to the wrong semiclassical behavior. For completeness, the phenomenological dynamics in the $\mu_o$-scheme is presented in \appref{sec:muzero dynamics}.} and should be improved by a more sophisticated construction ($\mubar$-scheme) in which the value of discreteness parameters depends adaptively on the scale factors (e.g., $\mubar\propto1/\sqrt{{p}}$ is used in \cite{Ashtekar:2006wn}) and thus implements the underlying physics of quantum geometry of LQG more directly \cite{Ashtekar:2006wn,Chiou:2007mg}.

For the case with Kantowski-Sachs symmetry, there is a variety of possibilities to implement the $\mubar$-scheme discreteness. Two well-motivated constructions (referred to as the ``$\mubar$-scheme'' and ``$\mubar'$-scheme'') are focused on in this paper:
\begin{itemize}
\item
$\mubar$-scheme:
\be\label{eqn:mubar}
\mubar_b=\sqrt{\frac{\Delta}{{p_b}}}\,,\qquad
\mubar_c=\sqrt{\frac{\Delta}{{p_c}}}\,,
\ee
\item
$\mubar'$-scheme:
\be\label{eqn:mubar'}
\mubar'_b=\sqrt{\frac{\Delta}{p_c}}\,,\qquad
\mubar'_c=\frac{\sqrt{p_c\Delta}}{p_b}\,.
\ee
\end{itemize}
Here $\Delta$ is the \emph{area gap} in the full theory of LQG and $\Delta=2\sqrt{3}\pi\gamma\Pl^2$ for the standard choice (but other choices are also possible) with $\Pl=\sqrt{G\hbar}$ being the Planck length.

Either scheme has its own merits and until more detailed physics is investigated it remains arguable which one is more sensible. In particular, the $\mubar$-scheme (in the version for the Bianchi I model) is suggested in \cite{Chiou:2006qq}, since in the construction of the fundamental loop quantum theory the Hamiltonian constraint in the $\mubar$-scheme gives a difference equation in terms of affine variables and therefore the well-developed framework of the spatially flat and isotropic LQC can be straightforwardly adopted. However, it is argued in \cite{Bojowald:2007ra} that the $\mubar$-scheme may lead to an unstable difference equation. On the other hand, the $\mubar'$-scheme does not admit the desirable affine variables but it has the virtue over the $\mubar$-scheme that its phenomenological dynamics is independent of the choice of $\mathcal{I}$ as will be seen (although this virtue is not necessarily required when quantum corrections are taken into account). To explore their virtues and ramifications, we study both the $\mubar$-scheme and the $\mubar'$-scheme at the level of phenomenological dynamics in \secref{sec:mubar dynamics} and \secref{sec:mubar' dynamics}, respectively. (Motivations for both schemes and more comments on them can be found in Appendix B of \cite{Chiou:2008eg}.)

Before going into detail, we can get an idea where the quantum corrections become appreciable by estimating the quantities $\mubar_bb$, $\mubar_cc$, etc., which indicate how significant the quantum corrections are (quantum corrections are negligible if $\mubar_bb,\,\mubar_cc,\,\text{etc.}\,\ll 1$). Plugging the classical solutions \eqnref{eqn:cl sol of Kb}, \eqnref{eqn:cl sol of pc} and \eqnref{eqn:cl sol of pb} into \eqnref{eqn:mubar} and \eqnref{eqn:mubar'}, we have
\ba
\mubar_bb &=&\gamma\mubar_b \frac{K_b}{p_b}
=\gamma\Delta^{1/2}\frac{K_b}{p_b^{3/2}}
\rightarrow
\left\{
\begin{array}{lcl}
\infty & & \quad\text{as }t'\rightarrow\infty,\\
\,0 & & \quad\text{as }t'\rightarrow-\infty,\\
\end{array}
\right.\\
\mubar_cc &=&\gamma\mubar_c \frac{K_c}{p_c}
=\gamma\Delta^{1/2}\frac{K_c}{p_c^{3/2}}
\rightarrow
\left\{
\begin{array}{lcl}
\ \infty & & \quad\text{as }t'\rightarrow\infty,\\
\frac{\gamma\Delta^{1/2}K_c}{8G^3M^3} & & \quad\text{as }t'\rightarrow-\infty,\\
\end{array}
\right.
\ea
and
\ba
\mubar'_bb &=&\gamma\mubar'_b \frac{K_b}{p_b}
=4\pi\gamma\Delta^{1/2}\frac{K_b}{\bf V}
\rightarrow
\left\{
\begin{array}{lcl}
\infty & & \quad\text{as }t'\rightarrow\infty,\\
\,0 & & \quad\text{as }t'\rightarrow-\infty,\\
\end{array}
\right.\\
\mubar'_cc &=&\gamma\mubar'_c \frac{K_c}{p_c}
=4\pi\gamma\Delta^{1/2}\frac{K_c}{\bf V}
\rightarrow
\left\{
\begin{array}{lcl}
\infty & & \quad\text{as }t'\rightarrow\infty,\\
\infty & & \quad\text{as }t'\rightarrow-\infty.\\
\end{array}
\right.
\ea
Therefore, in the $\mubar$-scheme, the quantum corrections are significant near the classical singularity and negligible on the horizon provided that
\be\label{eqn:qm reg}
K_c\ll\frac{8G^3M^3}{\gamma\sqrt{\Delta}},
\ee
which can always be satisfied if we choose $\mathcal{I}$ small enough for a given $M$. On the other hand, in the $\mubar'$-scheme, both the classical singularity and the horizon receive quantum corrections.

\subsection{Phenomenological dynamics in the $\mubar$-scheme}\label{sec:mubar dynamics}
The phenomenological dynamics in the $\mubar$-scheme is specified by the Hamiltonian \eqnref{eqn:qm Hamiltonian} with $\mubar_b$, $\mubar_c$ given by \eqnref{eqn:mubar}.
At the level of phenomenological dynamics, the equations of motion are governed by the Hamilton's equations and the constraint that the Hamiltonian must vanish; these are
\ba
\label{eqn:qm eom 3}
\frac{dc}{dt'}&=&\{c,H'_\mubar\}=2G\gamma\,\frac{\partial\, H'_\mubar}{\partial p_c}
=-2\gamma^{-1}\left[\frac{3\sin(\mubar_cc)}{2\mubar_c}-\frac{c\cos(\mubar_cc)}{2}\right]
\left[\frac{\sin(\mubar_bb)}{\mubar_b}p_b\right],\\
\label{eqn:qm eom 4}
\frac{dp_c}{dt'}&=&\{p_c,H'_\mubar\}=-2G\gamma\,
\frac{\partial\, H'_\mubar}{\partial c}
=2\gamma^{-1}p_c\cos(\mubar_cc)
\left[\frac{\sin(\mubar_bb)}{\mubar_b}p_b\right],\\
\label{eqn:qm eom 5}
\frac{db}{dt'}&=&\{b,H'_\mubar\}=G\gamma\,\frac{\partial\, H'_\mubar}{\partial p_b}
=-\gamma^{-1}\left[\frac{3\sin(\mubar_bb)}{2\mubar_b}-\frac{b\cos(\mubar_bb)}{2}\right]
\left[\frac{\sin(\mubar_bb)}{\mubar_b}p_b+\frac{\sin(\mubar_cc)}{\mubar_c}p_c\right]-\gamma p_b,\\
\label{eqn:qm eom 6}
\frac{dp_b}{dt'}&=&\{p_b,H'_\mubar\}=-G\gamma\,
\frac{\partial\, H'_\mubar}{\partial b}
=\gamma^{-1}p_b\cos(\mubar_bb)
\left[\frac{\sin(\mubar_bb)}{\mubar_b}p_b+\frac{\sin(\mubar_cc)}{\mubar_c}p_c\right],
\ea
as well as
\be\label{eqn:qm eom 7}
H'_\mubar=0
\quad\Rightarrow\quad
2\frac{\sin(\mubar_bb)}{\mubar_b}\frac{\sin(\mubar_cc)}{\mubar_c}\,p_bp_c
+\left[\left(\frac{\sin(\mubar_bb)}{\mubar_b}\right)^2+\gamma^2\right]p_b^2=0.
\ee
[Note that in the classical limit $\mubar_bb,\mubar_cc\rightarrow0$, we have
$\sin(\mubar_bb)/\mubar_b\rightarrow b$, $\sin(\mubar_cc)/\mubar_c\rightarrow c$ and
$\cos(\mubar_bb),\,\cos(\mubar_cc)\rightarrow1$. By inspection, it follows that
\eqnref{eqn:qm eom 3}--\eqnref{eqn:qm eom 7} reduce to their
classical counterparts \eqnref{eqn:cl eom 3}--\eqnref{eqn:cl eom 7} in the classical limit.]
Also notice that \eqnref{eqn:qm eom 4} and \eqnref{eqn:qm eom 6} lead to
\ba
\label{eqn:qm b}
\frac{\sin(\mubar_bb)}{\mubar_b}&=&\frac{1}{\cos(\mubar_cc)}
\frac{\gamma}{2p_c^{1/2}}\frac{dp_c}{d\tau},\\
\label{eqn:qm c}
\frac{\sin(\mubar_cc)}{\mubar_c}&=&\frac{1}{\cos(\mubar_bb)}
\frac{\gamma}{p_c^{1/2}}\frac{dp_b}{d\tau}
-\frac{1}{\cos(\mubar_cc)}\frac{\gamma p_b}{2p_c^{3/2}}\frac{dp_c}{d\tau},
\ea
which are the modifications of \eqnref{eqn:cl b} and \eqnref{eqn:cl c} with quantum corrections.

Combining \eqnref{eqn:qm eom 3} and \eqnref{eqn:qm eom 4}, we have
\be\label{eqn:qm dKc/dt'}
\left[\frac{3\sin(\mubar_cc)}{2\mubar_c}
-\frac{c\cos(\mubar_cc)}{2}\right]\frac{dp_c}{dt'}
+p_c\cos(\mubar_cc)\frac{dc}{dt'}
=\frac{d}{dt'}\left[p_c\frac{\sin(\mubar_cc)}{\mubar_c}\right]=0,
\ee
which, in accordance with the classical counterpart \eqnref{eqn:Kc}, yields the constant of motion:
\be\label{eqn:qm Kc}
p_c\frac{\sin(\mubar_cc)}{\mubar_c}
=\gamma K_c.
\ee
Similarly, \eqnref{eqn:qm eom 5} and \eqnref{eqn:qm eom 6} lead to
\be\label{eqn:qm dKb/dt'}
\left[\frac{3\sin(\mubar_bb)}{2\mubar_b}
-\frac{b\cos(\mubar_bb)}{2}\right]\frac{dp_b}{dt'}
+p_b\cos(\mubar_bb)\frac{db}{dt'}
=\frac{d}{dt'}\left[p_b\frac{\sin(\mubar_bb)}{\mubar_b}\right]
=-\gamma p_b^2\cos(\mubar_bb).
\ee
In accordance with the classical counterpart \eqnref{eqn:Kb}, we define
\be\label{eqn:qm Kb}
p_b\frac{\sin(\mubar_bb)}{\mubar_b}=:\gamma \bar{K}_b(t').
\ee
The Hamiltonian constraint \eqnref{eqn:qm eom 7} now read as
\be
2\bar{K}_bK_c+\bar{K}_b^2+p_b^2=0
\ee
and $\bar{K}_b$ satisfies the differential equation:
\be\label{eqn:qm diff eq for Kb}
\frac{d\bar{K}_b}{dt'}
=\cos(\mubar_bb)\left(2\bar{K}_bK_c+\bar{K}_b^2\right).
\ee
Substituting \eqnref{eqn:qm Kc} and \eqnref{eqn:qm Kb} into \eqnref{eqn:qm eom 4} and \eqnref{eqn:qm eom 6} yields
\ba
\label{eqn:qm diff eq for pc}
\frac{1}{p_c}\frac{dp_c}{dt'}
&\equiv& \frac{{\bf V}}{4\pi p_c}\frac{dp_c}{d\tau}=
2\cos(\mubar_cc)\bar{K}_b,\\
\label{eqn:qm diff eq for pb}
\frac{1}{p_b}\frac{dp_b}{dt'}
&\equiv& \frac{{\bf V}}{4\pi p_b}\frac{dp_b}{d\tau}=
\cos(\mubar_bb)\left[\bar{K}_b+K_c\right].
\ea
Note that, as in the classical dynamics, it follows from \eqnref{eqn:qm diff eq for Kb} that the flipping $K_c\longrightarrow-K_c$ gives rise to $\bar{K}_b(t')\longrightarrow-\bar{K}_b(-t')$ and thus corresponds to the time reversal according to \eqnref{eqn:qm diff eq for pc} and \eqnref{eqn:qm diff eq for pb}.

Equations \eqnref{eqn:qm diff eq for pc} and \eqnref{eqn:qm diff eq for pb} are the modifications of their classical counterparts \eqnref{eqn:cl diff eq for pc} and \eqnref{eqn:cl diff eq for pb}. Notice that the presence of the $\cos(\cdots)$ terms gives rise to the \emph{repulsive} behavior of gravity as the evolution departs from the classical solution. More precisely, in the $\mubar$-scheme phenomenological dynamics, $p_c$ and $p_b$ get bounced whenever $\cos(\mubar_cc)$ or $\cos(\mubar_bb)$ flips signs, respectively. To find out the exact moment of occurrence of the bounces, we investigate $\cos(\mubar_cc)$ and $\cos(\mubar_bb)$ in more detail.

By \eqnref{eqn:qm Kc} and \eqnref{eqn:qm Kb}, we have
\ba
\label{eqn:qm cosc}
\cos(\mubar_cc)&=&\pm\left[1-\sin^2\mubar_cc\right]^{1/2}
=\pm\left[1-\frac{\gamma^2K_c^2\Delta}{p_c^3}\right]^{1/2},\\
\label{eqn:qm cosb}
\cos(\mubar_bb)&=&\pm\left[1-\sin^2\mubar_bb\right]^{1/2}
=\pm\left[1-\frac{\gamma^2\bar{K}_b^2\Delta}{p_b^3}\right]^{1/2}.
\ea
Consequently, $p_c$ and $p_b$ get bounced, as $\cos(\mubar_cc)$ flips signs in \eqnref{eqn:qm diff eq for pc} and $\cos(\mubar_bb)$ flips signs in \eqnref{eqn:qm diff eq for pb}, respectively, whenever
\ba
\label{eqn:qm bounce for pc}
p_c&=&\left(\gamma^2K_c^2\Delta\right)^{1/3}\ll4G^2M^2,\\
\label{eqn:qm bounce for pb}
p_b&=&\left(\gamma^2\bar{K}_b^2\Delta\right)^{1/3}
\approx\left(4\gamma^2K_c^2\Delta\right)^{1/3}\ll4^{4/3}G^2M^2,
\ea
where we have used \eqnref{eqn:qm reg} and exploited the fact that $K_b\rightarrow-2K_c$ as the classical solution is close to the singularity.

\begin{figure}
\begin{picture}(500,200)(0,0)

\put(-71,-545)
{
\scalebox{1}{\includegraphics{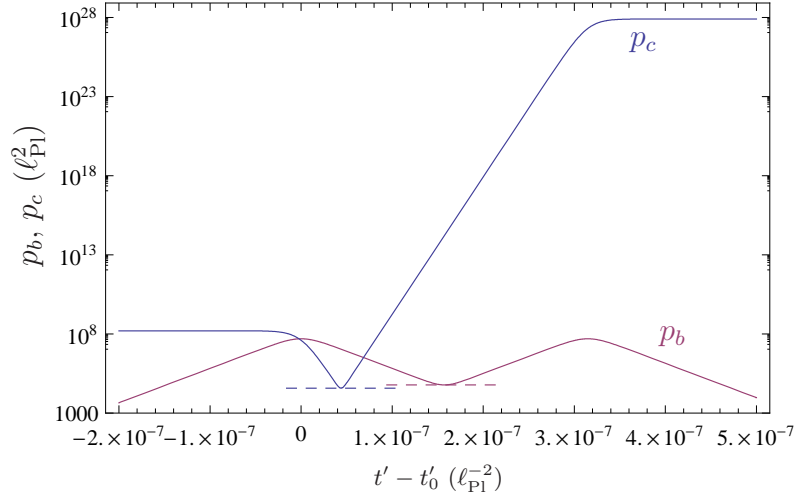}}
}

\end{picture}
\caption{\textbf{Solution in the $\mubar$-scheme phenomenological dynamics.} With the same initial condition as given in \figref{fig:classical solution} and the Barbero-Immirzi parameter is set to $\gamma=\ln2/(\sqrt{3}\pi)$. The conditions of \eqnref{eqn:qm bounce for pc} and \eqnref{eqn:qm bounce for pb} are indicated by dashed lines, at which $p_c$ and $p_b$ get bounced, respectively. The quantum bounce bridges the classical black hole (on the left side) with the classical white hole (on the right side). The asymptotic values of $p_c$ are different on both sides, indicating that the black hole mass $M$ and white hole mass $M'$ are different. On the other hand, $p_b$ is symmetric and, in particular, the peaks on both sides are of the same height, affirming that $K_c$ flips signs but its magnitude is unchanged as suggested in \eqnref{eqn:qm conjoined Kc}.}\label{fig:mubar solution}
\end{figure}

To sum up, \emph{the classical black hole singularity is replaced by the quantum bounce}, which makes both $p_c$ and $p_b$ bounced at the (different) epochs when the conditions \eqnref{eqn:qm bounce for pc} and \eqnref{eqn:qm bounce for pb} are met, respectively.
[Furthermore, with the $\cos(\mubar_bb)$ term in \eqnref{eqn:qm diff eq for Kb}, $\bar{K}_b$ becomes flat ($d\bar{K}_b/dt'=0$) exactly at the same time when $p_b$ gets bounced.] Across the quantum bounce, the evolution tends to be classical again [as $\cos(\mubar_bb),\ \cos(\mubar_cc) \rightarrow-1$ eventually]; as a result, the classical solution is connected with another classical solution through the quantum bounce.

Notice that the constant $K_c$ remains the same throughout the evolution. However, this does not mean that the parameter used to parametrize the classical evolutions on both sides of the bounce remains unchanged, since the physical meanings of $p_bb$ and $p_cc$ are changed before and after the bounce according to \eqnref{eqn:qm b} and \eqnref{eqn:qm c}.
In order to characterize the classical behaviors of the evolution in different classical periods, we define the ``$\text{\it effective }K_c$'' as
\ba\label{eqn:qm effective Kc}
\text{\it effective }K_c&:=&\gamma^{-1}p_c \left(\frac{\gamma}{p_c^{1/2}}\frac{dp_b}{d\tau}-\frac{\gamma p_b}{2p_c^{3/2}}\frac{dp_c}{d\tau}\right)\nn\\
&=&\gamma^{-1}\cos(\mubar_bb)\,p_c\frac{\sin(\mubar_cc)}{\mubar_c}
+\gamma^{-1}\left[\cos(\mubar_bb)-\cos(\mubar_cc)\right]
p_b\frac{\sin(\mubar_bb)}{\mubar_b}\nn\\
&=&\cos(\mubar_bb)K_c+\left[\cos(\mubar_bb)-\cos(\mubar_cc)\right]\bar{K}_b
\ea
and similarly the ``$\text{\it effective }K_b$'' as
\be\label{eqn:qm effective Kb}
\text{\it effective }K_b
:=\gamma^{-1}p_b\left(\frac{\gamma}{2p_c^{1/2}}\frac{dp_c}{d\tau}\right)
=\gamma^{-1}\cos(\mubar_cc) p_b \frac{\sin(\mubar_bb)}{\mubar_b}
=\cos(\mubar_cc)\bar{K}_b.\quad
\ee
In the classical regimes, $\cos(\mubar_bb)\approx\cos(\mubar_cc)\approx\pm1$ and we have
\ba
\label{eqn:qm conjoined Kc}
\text{\it effective }K_c&\approx&\pm K_c,\\
\label{eqn:qm conjoined Kb}
\text{\it effective }K_b&\approx&\pm K_b.
\ea
That is, on the other side of the bounce, both (classical) $K_c$ and $K_b$ flip signs and consequently \emph{the quantum bounce bridges the interior of a classical black hole with that of a classical white hole and vice versa}.

To know the mass of the white hole, in accordance with \eqnref{eqn:G2M2'}, we define the ``$\text{\it effective mass}$'' as
\be\label{eqn:qm effective mass}
\text{\it effective mass}:=
G^{-1}\frac{\sqrt{p_c}}{p_b}\
\abs{\text{\it effective }K_c}\abs{\text{\it effective }K_b}.
\ee
Note that $\text{\it effective }K_b$ approaches $\mp2K_c$ right before and after the bounce, leaving the factor $\abs{\text{\it effective }K_b}\abs{\text{\it effective }K_c}$ unchanged; however, the ratio $\sqrt{p_c}/p_b$ is not fixed and thus \eqnref{eqn:qm effective mass} yields unequal masses before and after the quantum bounce. Therefore, for a given black hole mass $M$, generally, the mass of the conjoined white hole (denoted as $M'$) is different from $M$. The exact value of $M'$ depends on the detail of the initial condition, which involves the choice of $\mathcal{I}$.

It is noteworthy that, in contrast to the classical dynamics, the phenomenological dynamics in the $\mubar$-scheme is \emph{dependent} on the choice of the finite sized interval $\mathcal{I}$. In particular, $M'$ depends on $\mathcal{I}$; moreover, \eqnref{eqn:qm bounce for pc} and \eqnref{eqn:qm bounce for pb}, which indicate occurrence of the bounce, are not invariant under rescaling of $\mathcal{I}$ (recall $p_b\propto L$, $p_c\propto L^0$ and $K_c\propto L$). For this matter, one might think that the $\mubar$-scheme quantization is simply ill-defined and should be discarded. However, it would be premature to dismiss the $\mubar$-scheme immediately as it is a common phenomenon that a quantum system reacts to macroscopic scales introduced by boundary conditions (for instance, the well-known ``conformal anomaly'' as a ``soft'' breaking of conformal symmetry). From the perspective of the full theory of LQG, the inhomogeneous degrees of freedom, which have been ignored in the symmetry-reduced minisuperspace formulation, could give rise to a macroscopic scale and thus account for the dependence on $\mathcal{I}$. (In the lattice refining model of \cite{Bojowald:2007ra}, this is indeed the case that, depending on the details of the refining procedure, the characteristic size of the lattice may leave imprints on the coarse-grained homogeneous description.) This suggests that the choice of $\mathcal{I}$ is not merely a gauge fixing but reflects the underlying physics of quantum inhomogeneity and thus has a physical consequence. In the language of the no-hair theorem, this physical consequence dictates that one extra parameter $M'$ (or equivalently, say $K_c$) is needed to completely characterize the extended Schwarzschild black hole, even though the information of $M'$ is hidden by the horizon and inaccessible (at least semiclassically) to the external observer. (cf. the apparent problem of dependence on $\mathcal{I}$ is absent in the phenomenological dynamics of the $\mubar$-scheme as will be seen in \secref{sec:mubar' dynamics}.)

For given initial conditions, the equations of motion can be solved numerically.\footnote{\label{footnote:numerical method}The common numerical methods (e.g., Runge-Kutta method) encounter numerical instability at some point if we directly solve the coupled differential equations \eqnref{eqn:qm eom 3}--\eqnref{eqn:qm eom 6}. To bypass this problem, which is only a numerical artifact, we solve the reduced coupled equations: \eqnref{eqn:qm diff eq for Kb}, \eqnref{eqn:qm diff eq for pc} and \eqnref{eqn:qm diff eq for pb} for three variables: $\bar{K}_b$, $p_c$ and $p_b$. The variables $b$ and $c$ can be obtained afterwards via \eqnref{eqn:qm Kc} and \eqnref{eqn:qm Kb}.}
The numerical solution is depicted in \figref{fig:mubar solution}.
Note that the bounces of $p_c$ and $p_b$ occur at the moments exactly as indicated in \eqnref{eqn:qm bounce for pc} and \eqnref{eqn:qm bounce for pb}. Also notice that $p_b$ is perfectly symmetric about the bounce, since \eqnref{eqn:qm diff eq for Kb} and \eqnref{eqn:qm diff eq for pb} are independent of $p_c$ and $c$ and, as a result, the evolution of $p_b$ is unaffected by the varying of $p_c$ (but not vice versa).

\subsection{Phenomenological dynamics in the $\mubar'$-scheme}\label{sec:mubar' dynamics}
The phenomenological dynamics in the $\mubar'$-scheme is specified by the effective Hamiltonian \eqnref{eqn:qm Hamiltonian} with $\mubar_b$, $\mubar_c$ replaced by $\mubar'_b$, $\mubar'_c$ given in \eqnref{eqn:mubar'}. To simplify the equations of motion, we choose a different lapse function $N=(p_b\sqrt{p_c})^{-1}$ associated with the new time variable $dt''=p_b\sqrt{p_c}\,d\tau$. With the new lapse, the Hamiltonian \eqnref{eqn:qm Hamiltonian} is further rescaled to the simpler form:
\ba\label{eqn:qm' Hamiltonian}
H''_{\mubar'}=
-\frac{1}{2G \gamma^2 \Delta}
\left\{
2\sin(\mubar'_bb)\sin(\mubar'_cc)+\sin^2(\mubar_bb)
+\Delta\frac{\gamma^2}{p_c}
\right\}.
\ea
Because $\abs{\sin(\mubar'_Ic_I)}\leq 1$, the vanishing of the Hamiltonian constraint $H''_{\mubar'}=0$ immediately implies
\be\label{eqn:qm' pc lower bound}
\abs{p_c}=\frac{\gamma^2\Delta}
{\abs{2\sin(\mubar'_bb)\sin(\mubar'_cc)+\sin^2(\mubar_bb)}}
\geq\frac{\gamma^2}{3}\Delta.
\ee
This suggests that $p_c$ is bounded below.

To know the detailed dynamics for each individual $p_b$ and $p_c$, in addition to the Hamiltonian constraint, we study the Hamilton's equations:
\ba
\label{eqn:qm' eom 3}
\frac{dc}{dt''}&=&\{c,H''_{\mubar'}\}=2 G\gamma\,\frac{\partial\, H''_{\mubar'}}{\partial p_c}\\
&=&
-\frac{c\mubar'_c\cos(\mubar'_cc)\sin(\mubar'_bb)}{\gamma\Delta\,p_c}
+\frac{b\mubar'_b\cos(\mubar'_bb)
\left[\sin(\mubar'_bb)+\sin(\mubar'_cc)\right]}{\gamma\Delta\, p_c}
+\frac{\gamma}{p_c^2},\nn\\
\label{eqn:qm' eom 4}
\frac{dp_c}{dt''}&=&\{p_c,H''_{\mubar'}\}=-2G\gamma\,\frac{\partial\, H''_{\mubar'}}{\partial c}=
\frac{2\mubar'_c\cos(\mubar'_cc)\sin(\mubar'_bb)}{\gamma\Delta},\\
\label{eqn:qm' eom 5}
\frac{db}{dt''}&=&\{b,H''_{\mubar'}\}=G\gamma\,\frac{\partial\, H''_{\mubar'}}{\partial p_b}=
\frac{c\mubar'_c\cos(\mubar'_cc)\sin(\mubar'_bb)}{\gamma\Delta\,p_b},\\
\label{eqn:qm' eom 6}
\frac{dp_b}{dt''}&=&\{p_b,H''_{\mubar'}\}=-G\gamma\,\frac{\partial\, H''_{\mubar'}}{\partial b}=
\frac{\mubar'_b\cos(\mubar'_bb)
\left[\sin(\mubar'_bb)+\sin(\mubar'_cc)\right]}{\gamma\Delta}.
\ea
Note that \eqnref{eqn:qm' eom 4} and \eqnref{eqn:qm' eom 6} give us
\ba
\label{eqn:qm' b}
\frac{\sin(\mubar'_bb)}{\mubar'_b}&=&\frac{1}{\cos(\mubar'_cc)}
\frac{\gamma}{2p_c^{1/2}}\frac{dp_c}{d\tau},\\
\label{eqn:qm' c}
\frac{\sin(\mubar'_cc)}{\mubar'_c}&=&\frac{1}{\cos(\mubar'_bb)}
\frac{\gamma}{p_c^{1/2}}\frac{dp_b}{d\tau}
-\frac{1}{\cos(\mubar'_cc)}\frac{\gamma p_b}{2p_c^{3/2}}\frac{dp_c}{d\tau},
\ea
which are the modifications of \eqnref{eqn:cl b} and \eqnref{eqn:cl c} with quantum corrections.

Inspecting \eqnref{eqn:qm' eom 3}--\eqnref{eqn:qm' eom 6}, we have
\be
\frac{d}{dt''}\left(p_cc-p_bb\right)=\frac{\gamma}{p_c}.
\ee
In accordance with the constant $K_c$ and the function $K_b(t')$ used for classical solutions in \eqnref{eqn:Kc} and \eqnref{eqn:Kb}, introducing the time-varying function $f(t'')$, we set
\be\label{eqn:qm' Kc}
p_cc=\gamma\left(K_c+f(t'')\right)
\ee
and
\be\label{eqn:qm' Kb}
p_bb=:\gamma\left(\bar{K}'_b(t'')+f(t'')\right),
\ee
where $\bar{K}'_b$ satisfies
\be\label{eqn:qm' diff eq for Kb}
p_b^2p_c\frac{d\bar{K}'_b}{dt''}=
\frac{d\bar{K}'_b}{dt'}=-p_b^2,
\ee
which is to be compared with the classical counterpart \eqnref{eqn:Kb}.
Starting in a classical regime, we set $K_c\approx\gamma^{-1}p_cc$ and $f\approx0$.

Substituting \eqnref{eqn:qm' Kc} and \eqnref{eqn:qm' Kb} into \eqnref{eqn:qm' Hamiltonian}, we have the complicated expression for the Hamiltonian constraint $H''_{\mubar'}=0$:
\be\label{eqn:qm' eom 7}
2\sin\left(\sqrt{\frac{\Delta\gamma^2}{p_b^2p_c}}\,(\bar{K}'_b+f)\right)
\sin\left(\sqrt{\frac{\Delta\gamma^2}{p_b^2p_c}}\,(K_c+f)\right)
+\sin^2\left(\sqrt{\frac{\Delta\gamma^2}{p_b^2p_c}}\,(\bar{K}'_b+f)\right)
=-\frac{\Delta\gamma^2}{p_c},
\ee
which reduces to
\be\label{eqn:qm' Kb Kc and pb}
2\bar{K}'_b K_c+\bar{K}_b^{\prime\,2}+p_b^2=0.
\ee
in the classical limit as $p_b^2p_c\gg\Delta\gamma^2K_\phi^2\sim \Delta\gamma^2K_c^2\sim \Delta\gamma^2K_{b,\pm}^2$ and with $f\approx0$.

As discussed earlier, in the $\mubar$-scheme, the quantum corrections take effect both near the classical singularity and the event horizon. Thus, it is expected that the classical singularity is resolved and replaced by the late time quantum bounce and the event horizon is diffused by the early time quantum bounce. Across the quantum bounces, we would guess, the evolution becomes classical again; as a result, the late/early time quantum bounce bridges one cycle of classical evolution with the next/previous classical cycle.

As in the $\mubar$-scheme, the constant $K_c$ remains fixed throughout the evolution but since $p_bb$ and $p_cc$ have different physical meanings before and after the quantum bounce according to \eqnref{eqn:qm' b} and \eqnref{eqn:qm' c}, analogous to \eqnref{eqn:qm effective Kc}, \eqnref{eqn:qm effective Kb} and \eqnref{eqn:qm effective mass}, we define
\ba\label{eqn:qm' effective Kc}
\text{\it effective }K_c&:=&\gamma^{-1}p_c \left(\frac{\gamma}{p_c^{1/2}}\frac{dp_b}{d\tau}-\frac{\gamma p_b}{2p_c^{3/2}}\frac{dp_c}{d\tau}\right)\nn\\
&=&\gamma^{-1}\cos(\mubar'_bb)\,p_c\frac{\sin(\mubar'_cc)}{\mubar'_c}
+\gamma^{-1}\left[\cos(\mubar'_bb)
-\cos(\mubar'_cc)\right]p_b\frac{\sin(\mubar'_bb)}{\mubar'_b},\\
\label{eqn:qm' effective Kb}
\text{\it effective }K_b
&:=&\gamma^{-1}p_b\left(\frac{\gamma}{2p_c^{1/2}}\frac{dp_c}{d\tau}\right)
=\gamma^{-1}\cos(\mubar'_cc) p_b \frac{\sin(\mubar'_bb)}{\mubar'_b}
\qquad\qquad\qquad\quad
\ea
and
\be\label{eqn:qm' effective mass}
\text{\it effective mass}:=
G^{-1}\frac{\sqrt{p_c}}{p_b}\
\abs{\text{\it effective }K_c}\abs{\text{\it effective }K_b}
\ee
to characterize the classical evolution in different classical periods.

Starting with $f\approx0$ and $p_cc\approx \gamma K_c$ in a given cycle of classical phase, we would guess $f$ varies widely when it undergoes the bounce but anchors to a nonzero constant (such that $\text{\it effective }K_c\approx K_c+f$) in the consecutive classical cycle when it jumps over the bounce. (The numerical analysis shows that this indeed is the case.) At the epoch right before the late time bounce, we have $\text{\it effective }K_c\approx K_c$, $\text{\it effective }K_b\approx K_b\approx-2K_c$ and $p_b\approx0$; immediately after the late time bounce, we then have $\text{\it effective }K_c\approx K_c+f$, $\text{\it effective }K_b\approx-2K_c+f$ and $p_b\approx0$, which should satisfy the classical Hamiltonian constraint \eqnref{eqn:Kb Kc and pb} and thus give
\be\label{eqn:qm' Kc and f}
2(-2K_c+f)(K_c+f)+(-2K_c+f)^2\approx0.
\ee
This yields $f\approx0$ or $f\approx2K_c$ and consequently implies that the $\text{\it effective }K_c$ is altered to be $K_c+f\approx3K_c$ and $\text{\it effective }K_b\approx K_b+f\approx0$ right after the late time bounce. Similarly, starting with $\text{\it effective }K_c\approx K_c$, $\text{\it effective }K_b\approx K_b\approx0$ and $p_b\approx0$ at the epoch close to the early time bounce, we can infer that $f\approx-2K_c/3$, $\text{\it effective }K_c\approx K_c+f\approx K_c/3$ and $\text{\it effective }K_b\approx K_b+f\approx-2K_c/3$ immediately across the early time bounce. We then conclude that \emph{the quantum bounce resolves the black hole singularity and bridges it with the diffused horizon of another black hole} (not white hole!); the parameter $K_c$ in one cycle of classical phase is shifted to $3K_c$ in the next classical cycle and to $K_c/3$ in the previous cycle. Schematically, the varying of the $\text{\it effective }K_c$ is summarized as
\be\label{eqn:qm' conjoined Kc}
\cdots\quad
\frac{K_c}{3^2}
\;
{
\underleftrightarrow{\ \mbox{\tiny bounce}\ }\\
\atop
\mbox{}
}
\;
\frac{K_c}{3}
\;
{
\underleftrightarrow{\ \mbox{\tiny bounce}\ }\\
\atop
\mbox{}
}
\;
K_c
\;
{
\underleftrightarrow{\ \mbox{\tiny bounce}\ }\\
\atop
\mbox{}
}
\;
3K_c
\;
{
\underleftrightarrow{\ \mbox{\tiny bounce}\ }\\
\atop
\mbox{}
}
\;
3^2K_c
\quad\cdots.
\ee

To find out the precise condition for the occurrence of quantum bounces, by substituting \eqnref{eqn:qm' Kc} and \eqnref{eqn:qm' Kb} into \eqnref{eqn:qm' eom 4} and \eqnref{eqn:qm' eom 6}, we study the differential equations:
\ba
\label{eqn:qm' diff eq for pc}
\frac{1}{p_c}\frac{dp_c}{dt'}
&\equiv& \frac{{\bf V}}{4\pi p_c}\frac{dp_c}{d\tau}=
2\sqrt{\frac{p_b^2p_c}{\gamma^2\Delta}}
\ \cos\left(\sqrt{\frac{\gamma^2\Delta}{p_b^2p_c}}\,(K_c+f)\right)
\sin\left(\sqrt{\frac{\gamma^2\Delta}{p_b^2p_c}}\,(\bar{K}_b'+f)\right),\\
\label{eqn:qm' diff eq for pb}
\frac{1}{p_b}\frac{dp_b}{dt'}
&\equiv& \frac{{\bf V}}{4\pi p_b}\frac{dp_b}{d\tau}=
\sqrt{\frac{p_b^2p_c}{\gamma^2\Delta}}
\ \cos\left(\sqrt{\frac{\gamma^2\Delta}{p_b^2p_c}}\,(\bar{K}_b'+f)\right)\nn\\
&&\quad\qquad\qquad\times\left[
\sin\left(\sqrt{\frac{\gamma^2\Delta}{p_b^2p_c}}\,(\bar{K}_b'+f)\right)
+\sin\left(\sqrt{\frac{\gamma^2\Delta}{p_b^2p_c}}\,(K_c+f)\right)
\right].
\ea
These are the modifications of the classical counterparts \eqnref{eqn:cl diff eq for pc} and \eqnref{eqn:cl diff eq for pb}.

Similar to the case of \eqnref{eqn:qm diff eq for pc} in the $\mubar$-scheme, $p_c$ gets bounced once the $\cos(\cdots)$ term in \eqnref{eqn:qm' diff eq for pc} flips signs.\footnote{The numerical result further shows that once the $\cos(\cdots)$ term in \eqnref{eqn:qm diff eq for pc} or \eqnref{eqn:qm diff eq for pb} flips signs from $+1$ to $-1$, it quickly flips back to $+1$. Both $\cos(\mubar'_cc)$ and $\cos(\mubar'_bb)$ flip \emph{twice} during the bouncing period. This concurs with the previous finding that the quantum bounce bridges the black hole with another black hole (instead of a white hole). By contrast, in the $\mubar$-scheme, $\cos(\mubar_cc)$ and $\cos(\mubar_cc)$ flip from $+1$ to $-1$ only once and thus the quantum bounce conjoins a black hole with a white hole.} This happens when
\be\label{eqn:qm' condition 1}
\cos\left(\sqrt{\frac{\gamma^2\Delta}{p_b^2p_c}}\,(K_c+f)\right)=0
\quad\Rightarrow\quad
K_c+f=\frac{\pi}{2}\sqrt{\frac{p_b^2p_c}{\gamma^2\Delta}}\,.
\ee
Assuming $p_b$ also gets bounced roughly around the same moment,\footnote{\label{footnote:bounces close}This is because \eqnref{eqn:qm' diff eq for pc} and \eqnref{eqn:qm' diff eq for pb} are coupled through ${\bf V}=4\pi p_b\sqrt{p_c}$. We can see that this is indeed the case in the numerical solution.} at which \eqnref{eqn:qm' condition 1} is satisfied, we have the approximation:
\ba\label{eqn:qm' condition 2}
&&\sin\left(\sqrt{\frac{\gamma^2\Delta}{p_b^2p_c}}\,(\bar{K}_b'+f)\right)
=\sin\left(\frac{\pi}{2}+\sqrt{\frac{\gamma^2\Delta}{p_b^2p_c}}\,
(\bar{K}_b'-K_c)\right)\nn\\
&=&\cos\left(\sqrt{\frac{\gamma^2\Delta}{p_b^2p_c}}\,(\bar{K}_b'-K_c)\right)
\approx 1-\frac{1}{2!}\frac{\gamma^2\Delta}{p_b^2p_c}\,(\bar{K}_b'-K_c)^2+
\frac{1}{4!}\left(\frac{\gamma^2\Delta}{p_b^2p_c}\right)^2(\bar{K}_b'-K_c)^4
+\cdots.\quad
\ea
Taking \eqnref{eqn:qm' condition 1} and \eqnref{eqn:qm' condition 2} into \eqnref{eqn:qm' eom 7}, we have
\ba
0 &\approx& 2\frac{p_b^2p_c}{\gamma^2\Delta}
\left[1-\frac{\gamma^2\Delta}{2p_b^2p_c}(\bar{K}_b'-K_c)^2
+\frac{1}{4!}\left(\frac{\gamma^2\Delta}{p_b^2p_c}\right)^2(\bar{K}_b'-K_c)^4\right]
\nn\\
&&\;+\frac{p_b^2p_c}{\gamma^2\Delta}
\left[1-\frac{\gamma^2\Delta}{2p_b^2p_c}(\bar{K}_b'-K_c)^2
+\frac{1}{4!}\left(\frac{\gamma^2\Delta}{p_b^2p_c}\right)^2(\bar{K}_b'-K_c)^4\right]^2
+p_b^2+\cdots,
\ea
which, provided that $p_b,\,p_c\gg\gamma^2\Delta$ when the bounce occurs,\footnote{\label{footnote:approx2}We will see that this is true until $p_c$ eventually descends into the deep Planck regime in the far late time.} leads to the condition for occurrence of the bounce in $p_c$:
\be\label{eqn:qm' bounce for pc}
\frac{\gamma^2\Delta}{p_b^2p_c}\approx
\frac{2(3-\sqrt{3}\,)}{(\bar{K}_b'-K_c)^2}
\approx
\left\{
\begin{array}{lcl}
\frac{2(3-\sqrt{3})}{9K_c^2} & & \quad\text{for the late time bounce in }p_c,\\
\frac{2(3-\sqrt{3})}{K_c^2} & & \quad\text{for the early time bounce in }p_c.
\end{array}
\right.
\ee
Here, we have exploited the fact that \eqnref{eqn:cl diff eq for Kb} and \eqnref{eqn:qm' diff eq for Kb} are formally identical and therefore $\bar{K}'_b$ remains almost constant ($\bar{K}'_b\approx K_b\rightarrow -2K_c\text{ or }0$) close to the late/early time bounce even when quantum corrections take effect later. (In the bouncing period, the quantum effect varies $f$ dramatically but modifies $\bar{K}'_b$ only slightly.)\footnote{Do not confuse $\bar{K}'_b$ with $\text{\it effective }K_b$. The former remains constant through the bounce while the latter is offset by $f$.}

Similarly, $p_b$ gets bounced once the $\cos(\cdots)$ term in \eqnref{eqn:qm' diff eq for pb} flips signs. Following the same argument above, we conclude that the big bounce of $p_b$ happens when
\be
0 \approx 2\frac{p_b^2p_c}{\gamma^2\Delta}
\left[1-\frac{\gamma^2\Delta}{2p_b^2p_c}(\bar{K}_b'-K_c)^2
+\frac{1}{4!}\left(\frac{\gamma^2\Delta}{p_b^2p_c}\right)^2(\bar{K}_b'-K_c)^4
\right]
+\frac{p_b^2p_c}{\gamma^2\Delta}
+p_b^2+\cdots,
\ee
which leads to the condition for occurrence of the bounce in $p_b$:
\be\label{eqn:qm' bounce for pb}
\frac{\gamma^2\Delta}{p_b^2p_c}\approx
\frac{6}{(\bar{K}_b'-K_c)^2}
\approx
\left\{
\begin{array}{lcl}
\frac{2}{3K_c^2} & & \quad\text{for the late time bounce in }p_b,\\
\frac{6}{K_c^2} & & \quad\text{for the early time bounce in }p_b.
\end{array}
\right.
\ee

Since the Taylor series of $\cos x=1-x^2/2+x^4/4!+\cdots$ converges very rapidly, the approximation made above is fairly accurate if
\be
\abs{x}=\sqrt{\frac{\gamma^2\Delta}{p_b^2p_c}}\,\abs{\bar{K}_b'-K_c}<\pi,
\ee
which is satisfied for both \eqnref{eqn:qm' bounce for pc} and \eqnref{eqn:qm' bounce for pb}.

Knowing the conditions for occurrence of bounces, we are able to estimate the black hole mass in different classical cycles. Let $M$ be the mass of a given classical cycle with the constant $K_c$; \eqnref{eqn:G2M2'} then tells us
\be\label{eqn:condition1}
G^2M^2=p_c(\check{t}''_-)
\left(\frac{K_b(\check{t}''_-)K_c}{p_b(\check{t}''_-)^2}\right)^2
\approx4\frac{p_c(\check{t}''_-)}{p_b(\check{t}''_-)^4}K_c^4,
\ee
where we denote the epoch of the late time bounce in $p_c$ as $\check{t}''$ and the instant right before $\check{t}''$ as $\check{t}''_-$, at which the quantum effect is still negligible and the evolution is classical enough so that $K_b(\check{t}''_-)\rightarrow-2K_c$. On the other hand, let $\mathfrak{M}$ be the mass of the black hole in the next classical cycle after the big bounce; \eqnref{eqn:G2M2} then gives us
\be\label{eqn:condition2}
G^2\mathfrak{M}^2=p_c(\check{t}''_+)
\left(\frac{K_c}{K_b(\check{t}''_+)+2K_c}\right)^2
\approx\frac{p_c(\check{t}''_+)}{4},
\ee
where we denote the instant right after $\check{t}''$ as $\check{t}''_+$, at which the evolution is classical enough and thus $K_b(\check{t}''_+)\rightarrow0$. Meanwhile, at $\check{t}''$, \eqnref{eqn:qm' bounce for pc} also tells us
\be\label{eqn:condition3}
\frac{\gamma^2\Delta}{p_b(\check{t}'')^2p_c(\check{t}'')}
\approx\frac{2(3-\sqrt{3}\,)}{9K_c^2}.
\ee
Assuming that $\check{t}$, $\check{t}_+$ and $\check{t}_-$ are fairly close to one another so that $p_c(\check{t}_+)\approx p_c(\check{t}_-)\approx p_c(\check{t})$ and $p_b(\check{t})\approx p_b(\check{t}_-)$, we can infer from \eqnref{eqn:condition1}, \eqnref{eqn:condition2} and \eqnref{eqn:condition3} that
\be
\mathfrak{M}(M)\approx\left(\frac{9}{32(3-\sqrt{3})}\right)^{1/3}
\left(\frac{\gamma^2\Delta M}{G^2}\right)^{1/3}
\simeq0.605\left(\frac{\gamma^2\Delta M}{G^2}\right)^{1/3}.
\ee
While this analysis gives a very good estimate with small error due to approximation, the detailed numerical solution gives the more precise result
\be\label{eqn:qm' precise effective mass}
\mathfrak{M}(M)\simeq0.524\left(\frac{\gamma^2\Delta M}{G^2}\right)^{1/3}
\simeq1.161\left(m_{\rm Pl}^2M\right)^{1/3},
\ee
where $m_{\rm Pl}:=\sqrt{\hbar/G}$ is the Planck mass.
As a result, the $\text{\it effective mass}$ is tremendously decreased by the late time bounces until it eventually approaches $\sim m_{\rm Pl}$; schematically, the varying of the $\text{\it effective mass}$ is summarized as
\be\label{eqn:qm' conjoined mass}
\cdots\quad
\mathfrak{M}^{-1}(\mathfrak{M}^{-1}(M))
\;
{
\underleftrightarrow{\ \mbox{\tiny bounce}\ }\\
\atop
\mbox{}
}
\;
\mathfrak{M}^{-1}(M)
\;
{
\underleftrightarrow{\ \mbox{\tiny bounce}\ }\\
\atop
\mbox{}
}
\;
M
\;
{
\underleftrightarrow{\ \mbox{\tiny bounce}\ }\\
\atop
\mbox{}
}
\;
\mathfrak{M}(M)
\;
{
\underleftrightarrow{\ \mbox{\tiny bounce}\ }\\
\atop
\mbox{}
}
\;
\mathfrak{M}(\mathfrak{M}(M))
\quad\cdots.
\ee

\begin{figure}
\begin{picture}(500,200)(0,0)

\put(-71,-546)
{
\scalebox{1}{\includegraphics{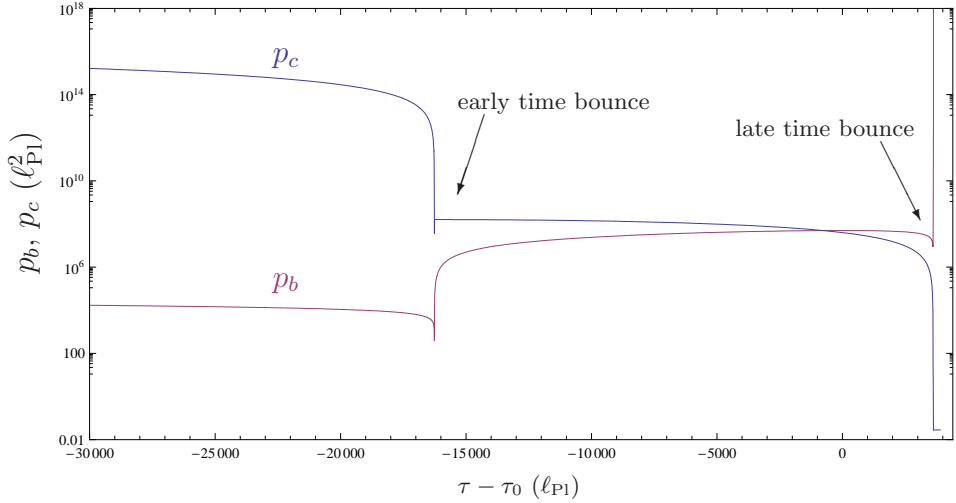}}
}

\end{picture}
\caption{\textbf{Solution in the $\mubar'$-scheme phenomenological dynamics.}  With the same initial condition as given in \figref{fig:classical solution} and the Barbero-Immirzi parameter is set to $\gamma=\ln2/(\sqrt{3}\pi)$. $\tau$ is the proper time and $\tau_0=\tau(t'_0)$. The event horizon is diffused by the early time bounce and connected to another black hole of a larger mass. The classical singularity is resolved and replaced by the late time bounce, which bridges the black hole with another black hole of a smaller mass. The vicinity of the early time bounce is zoomed in in \figref{fig:early time bounce} and that of the late time bounce in \figref{fig:late time bounce}.}\label{fig:mubar' solution}
\end{figure}

\begin{figure}
\begin{picture}(500,540)(0,0)

\put(-63,-202)
{
\scalebox{1}{\includegraphics{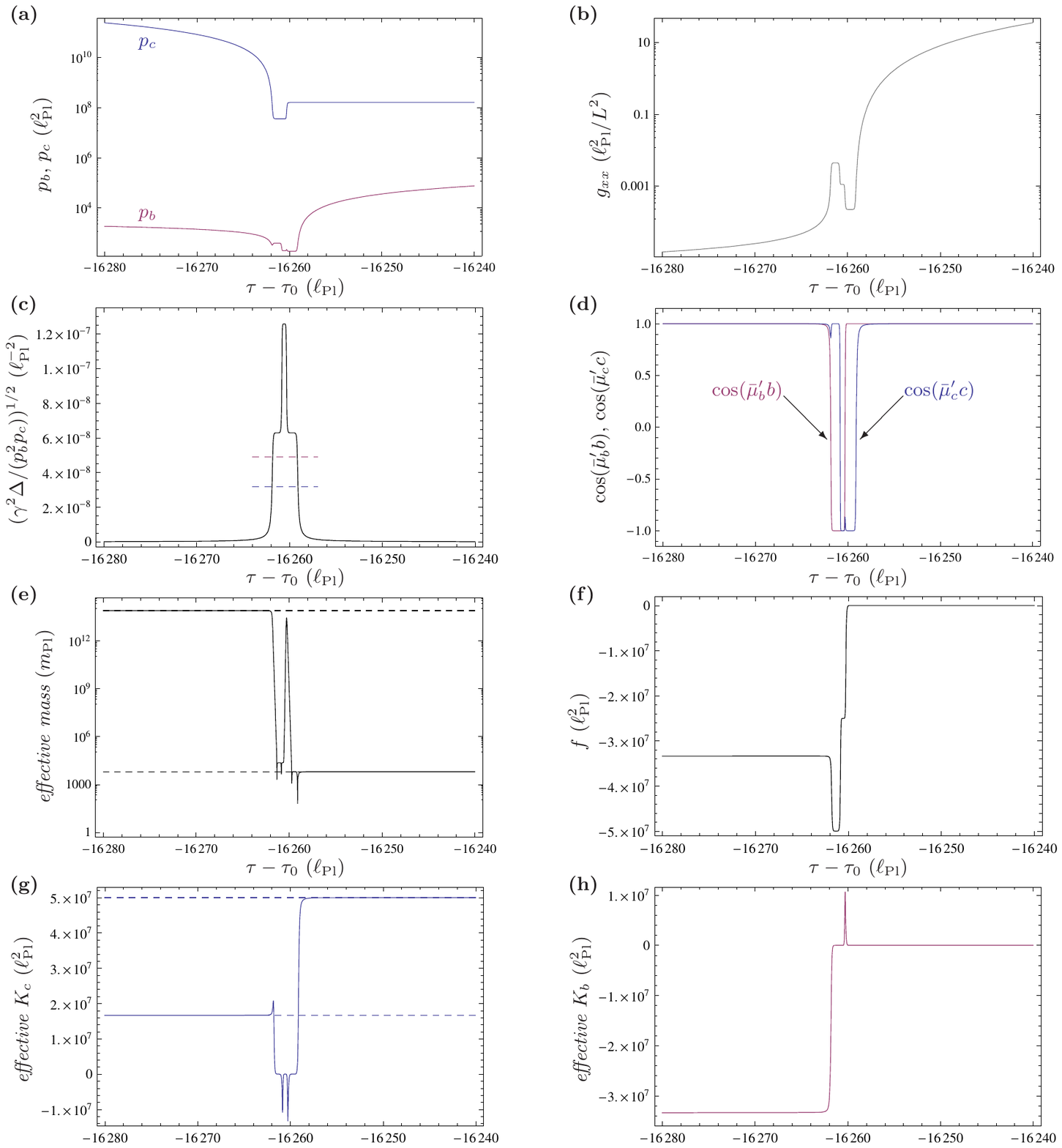}}
}

\end{picture}
\caption{\textbf{Zoom in of the early time bounce in \figref{fig:mubar' solution}.}
\textbf{(a)} $p_b(\tau)$ and $p_c(\tau)=g_{\Omega\Omega}(\tau)$. The epochs of bounces in $p_b$ and $p_c$ are very close to each other.
\textbf{(b)} $g_{xx}(\tau)$.
\textbf{(c)} $\sqrt{\gamma^2\Delta/(p_b^2p_c)}$, which signals the occurrence of bounces. The conditions of \eqnref{eqn:qm' bounce for pc} and \eqnref{eqn:qm' bounce for pb} are indicated by dashed lines.
\textbf{(d)} $\cos(\mubar'_bb)$ and $\cos(\mubar'_cc)$, fairly close to each other; both flip signs twice when undergoing the quantum bounce.
\textbf{(e)} $\text{\it effective mass}$, with the constants $M$ and $\mathfrak{M}^{-1}(M)$ indicated by dashed lines. See \eqnref{eqn:qm' conjoined mass}.
\textbf{(f)} $f(\tau)$. $f\approx0$ in the classical cycle on the right and $f\approx-2K_c/3$ on the left.
\textbf{(g)} $\text{\it effective } K_c$, with the constants $K_c$ and $K_c/3$ indicated by dashed lines.
\textbf{(h)} $\text{\it effective } K_b$, which becomes $0$ on the right of the bounce and $-2K_c/3$ on the left.
[For \textbf{(f), (g)} and \textbf{(h)}, see \eqnref{eqn:qm' conjoined Kc} and the text prior to it for the details.]\\ \\.}\label{fig:early time bounce}
\end{figure}

\begin{figure}
\begin{picture}(500,540)(0,0)

\put(-63,-202)
{
\scalebox{1}{\includegraphics{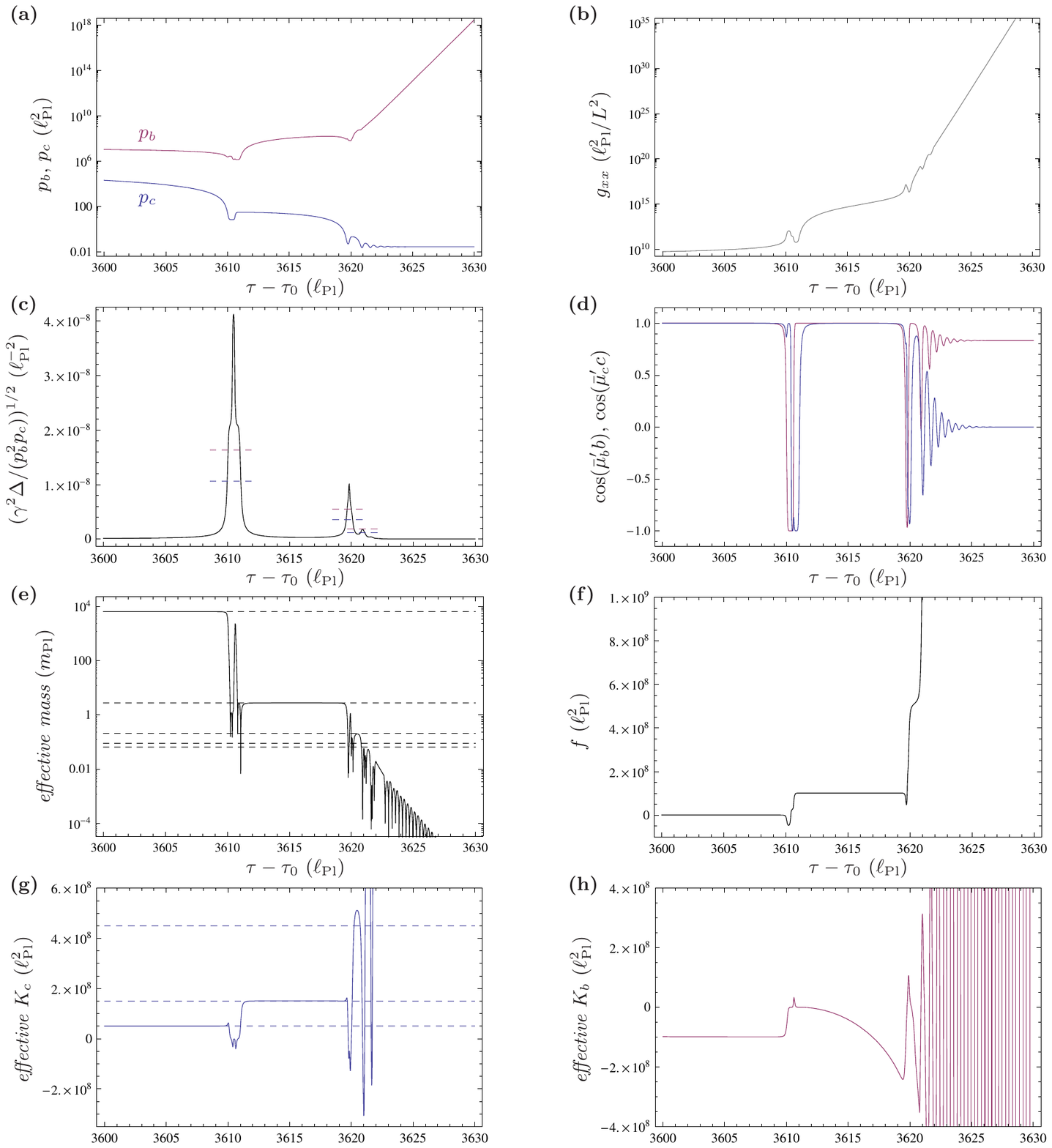}}
}

\end{picture}
\caption{\textbf{Zoom in of the late time bounce in \figref{fig:mubar' solution}.}
\textbf{(a)} $p_b(\tau)$ and $p_c(\tau)=g_{\Omega\Omega}(\tau)$. A few classical cycles are connected through quantum bounces. The semiclassicality of these cycles is however less and less established; eventually, $p_c$ descends into the deep Planck regime as $p_c\rightarrow\bar{p}_c$ and $p_b$ grows exponentially as $p_b\rightarrow\bar{p}_be^{\kappa\tau}$ with constants given by \eqnref{eqn:qm' pc asym} and \eqnref{eqn:qm' kappa asym}.
\textbf{(b)} $g_{xx}(\tau)$.
\textbf{(c)} $\sqrt{\gamma^2\Delta/(p_b^2p_c)}$, which signals the occurrence of bounces. The conditions of \eqnref{eqn:qm' bounce for pc} and \eqnref{eqn:qm' bounce for pb} with corresponding $\text{\it effective }K_c$ are indicated by dashed lines.
\textbf{(d)} $\cos(\mubar'_bb)$ and $\cos(\mubar'_cc)$, which are close to $+1$ in classical cycles but oscillate rapidly when the semiclassicality breaks down and eventually $\cos(\mubar'_cc)\rightarrow0$ and $\cos(\mubar'_bb)\rightarrow\cos\beta$ as given in \eqnref{eqn: qm' beta asym}.
\textbf{(e)} $\text{\it effective mass}$, with the constants $M$, $\mathfrak{M}(M)$, $\mathfrak{M}(\mathfrak{M}(M))$, \dots indicated by dashed lines. See \eqnref{eqn:qm' conjoined mass}.
\textbf{(f)} $f(\tau)$, which becomes constant in each classical cycle.
\textbf{(g)} $\text{\it effective } K_c$, with the constants $K_c$, $3K_c$ and $3^2K_c$ indicated by dashed lines.
\textbf{(h)} $\text{\it effective } K_b$.
[For \textbf{(f), (g)} and \textbf{(h)}, see \eqnref{eqn:qm' conjoined Kc} and the text prior to it for the details.]}\label{fig:late time bounce}
\end{figure}

The differential equations \eqnref{eqn:qm' eom 3}--\eqnref{eqn:qm' eom 6} can be solved numerically for a given initial condition. The numerical solution is shown in \figref{fig:mubar' solution}, which depicts both the late time and the early time quantum bounces. The vicinity of the early time bounce is zoomed in in \figref{fig:early time bounce} and that of the late time bounce in \figref{fig:late time bounce}. The early/late time quantum bounce bridges a classical phase with another classical phase in the previous/next cycle. Contrary to the $\mubar$-scheme, the epochs of bounces in $p_c$ and $p_b$ are very close to each other (see \footref{footnote:bounces close}). Toward the future, the $\text{\it effective } K_c$ becomes larger and larger while the $\text{\it effective mass}$ becomes smaller and smaller. As can be seen in \figref{fig:late time bounce}, the semiclassicality is less and less established and eventually $p_b$ grows exponentially (with respect to $\tau$) while $p_c$ asymptotically descends to a constant in the deep Planck regime, in which the quantum fluctuations become essential.

Although the validity of the semiclassical analysis might break down when the solution descends into the deep Planck regime,\footnote{In particular, the cotriad component $\omega_c=L\sqrt{g_{xx}}$ grows exponentially and the quantum corrections on it have to be taken into account.} it is still instructive to know the asymptotic behavior within the same phenomenological framework. To find out the asymptotic solution, we assume $p_c=\bar{p}_c$, $p_b=\bar{p}_b e^{\kappa \tau}$ with constants $\bar{p}_c$, $\bar{p}_b$ and $\kappa$. By \eqnref{eqn:qm' eom 4} and \eqnref{eqn:qm' eom 6}, we have
\ba
\frac{1}{p_c}\frac{dp_c}{d\tau}&=&0
=\frac{2}{\gamma\sqrt{\Delta}}\cos(\mubar'_cc)\sin(\mubar'_bb),\\
\frac{1}{p_b}\frac{dp_b}{d\tau}&=&\kappa
=\frac{1}{\gamma\sqrt{\Delta}}\cos(\mubar'_bb)
\left[\sin(\mubar'_bb)+\sin(\mubar'_cc)\right].
\ea
which yield $\mubar'_cc=(2n+1/2)\pi$ with $n\in\mathbb{Z}$ [such that $\cos(\mubar'_cc)=0$, $\sin(\mubar'_cc)$=1], $\mubar'_bb=\beta$ being a constant and consequently
\be\label{eqn:qm' asym 1}
\kappa=\frac{1}{\gamma\sqrt{\Delta}}\cos\beta\,(\sin\beta+1).
\ee
Substituting these into \eqnref{eqn:qm' eom 3} and \eqnref{eqn:qm' eom 5}, we have
\ba
\frac{db}{d\tau}&=&0
\quad\Rightarrow\; b=\bar{b}\text{ being a constant},\\
\label{eqn:qm' dc/dt}
\frac{dc}{d\tau}&=&
\bar{p}_b e^{\kappa\tau}
\left[\frac{\beta\cos\beta\,(\sin\beta+1)}{\gamma\Delta\sqrt{\bar{p}_c}}
+\frac{\gamma}{\bar{p}_c^{3/2}}\right]
\quad\Rightarrow\; c=\bar{c}\,e^{\kappa\tau}\text{ with a constant }\bar{c}.
\ea
Additionally, $(2n+1/2)\pi=\mubar'_cc=c\sqrt{\bar{p}_c\Delta}/p_b$ yields
\be\label{eqn:qm' bar c}
\bar{c}=\left(2n\pi+\frac{\pi}{2}\right)
\frac{\bar{p}_b}{\sqrt{\bar{p}_c\Delta}}.
\ee
Taking \eqnref{eqn:qm' bar c} into \eqnref{eqn:qm' dc/dt}, we have
\be\label{eqn:qm' asym 2}
\left(2n\pi+\frac{\pi}{2}\right)\kappa
=\frac{\beta\cos\beta\,(\sin\beta+1)}{\gamma\sqrt{\Delta}}
+\frac{\gamma\sqrt{\Delta}}{\bar{p}_c}.
\ee
Finally, the Hamiltonian constraint \eqnref{eqn:qm' Hamiltonian} reads as
\be\label{eqn:qm' asym 3}
2\sin\beta+\sin^2\beta+\frac{\Delta\gamma^2}{\bar{p}_c}=0.
\ee
Summing up \eqnref{eqn:qm' asym 1}, \eqnref{eqn:qm' asym 2} and \eqnref{eqn:qm' asym 3}, we have
\be
\left(2n\pi+\frac{\pi}{2}\right)\cos\beta\,(\sin\beta+1)
=\beta\cos\beta\,(\sin\beta+1)-2\sin\beta-\sin^2\beta,
\ee
the numerical solution of which is given by
\be\label{eqn: qm' beta asym}
\beta\simeq 2n\pi-0.587233,
\qquad
\cos\beta \simeq 0.832477.
\ee
By \eqnref{eqn:qm' asym 1} and \eqnref{eqn:qm' asym 3}, this leads to
\be\label{eqn:qm' pc asym}
\bar{p_c} \simeq 1.24823\,\gamma^2\Delta \simeq 0.0280788\,\Pl^2
\ee
and
\be\label{eqn:qm' kappa asym}
\kappa\simeq0.371235\,\gamma^{-1}\Delta^{-1/2}\simeq2.47518\,\Pl^{-1}.
\ee
These precisely agree with the asymptotic behaviors shown in \textbf{(a)} and \textbf{(d)} of \figref{fig:late time bounce}. Also note that \eqnref{eqn:qm' pc asym} is fairly close to the lower bound in \eqnref{eqn:qm' pc lower bound}.\footnote{It was claimed in \cite{Bohmer:2007wi} that the $\mubar$-scheme phenomenological dynamics extends a classical Schwarzschild black hole to a patch of a nonsingular charged Nariai universe, which gives constant $p_c$. However, a closer look suggests that the extended part is \emph{not} a patch of the classical Nariai universe but instead represents the quantum spacetime which \emph{formally} exhibits Nariai type metric, as the asymptotic constant $\bar{p}_c$ is in the deep Planck regime ($\alt\Pl^2$).}

Finally, as to the issue of dependence on $\mathcal{I}$, \eqnref{eqn:qm' b} and \eqnref{eqn:qm' c} imply that the quantities $\mubar'_bb$, $\mubar'_cc$ depend only on $p_b^{-1}dp_b/d\tau$, $p_c^{-1}dp_c/d\tau$ and thus are independent of $\mathcal{I}$ (recall $p_b\propto L$, $p_c\propto L^0$). Consequently, \eqnref{eqn:qm' Kc} and \eqnref{eqn:qm' Kb} tell us that $K_c$, $\bar{K}_b$ and $f$ all scale as $\propto L$. Therefore, the phenomenological dynamics given by \eqnref{eqn:qm' diff eq for pc} and \eqnref{eqn:qm' diff eq for pb} is \emph{completely independent} of the choice of $\mathcal{I}$ as is the classical dynamics. In particular, the choice of $\mathcal{I}$ has no effect on the conditions of bounce occurrence in \eqnref{eqn:qm' bounce for pc} and \eqnref{eqn:qm' bounce for pb}. This is a desirable feature that the $\mubar$-scheme does not have.
(However, if we further impose the quantum corrections on the eigenvalue of the cotriad operator $\hat{\omega}_c$, this invariance is broken again.)

\section{Scaling symmetry}\label{sec:scaling}

We have noted that the classical dynamics and the phenomenological dynamics in the $\mubar'$-scheme are both completely independent of the choice of the finite sized interval $\mathcal{I}$, whereas the phenomenological dynamics in the $\mubar$-scheme reacts to the physical size of $\mathcal{I}$. This can be rephrased in terms of the scaling symmetry;\footnote{A dynamical system is said to be invariant under a certain scaling if for a given solution [$p_b(\tau)$, $p_c(\tau)$, $b(\tau)$ and $c(\tau)$] to the dynamics, the rescaled functions also satisfy the equations of motion (i.e., Hamilton's equations and vanishing of Hamiltonian constraint). For the classical dynamics, the equations to be satisfied are \eqnref{eqn:cl eom 3}--\eqnref{eqn:cl eom 7}; for the $\mubar$-scheme, \eqnref{eqn:qm eom 3}--\eqnref{eqn:qm eom 7}; and for the $\mubar'$-scheme, \eqnref{eqn:qm' eom 3}--\eqnref{eqn:qm' eom 6} and \eqnref{eqn:qm' eom 7}.} that is, the classical dynamics and the $\mubar'$-scheme phenomenological dynamics are invariant under the following scaling:
\ba\label{eqn:symmetry 1}
p_b,\ p_c&\longrightarrow& l p_b,\ p_c,\nn\\
b,\ c&\longrightarrow& b,\ lc,\nn\\
K_c&\longrightarrow& lK_c,\nn\\
M&\longrightarrow& M.
\ea
(Note that the scaling for $K_c$ should be accompanied by the same scaling on $K_b$ in classical dynamics and on $\bar{K}'_b$ as well as $f$ in the $\mubar'$-scheme; that is $K_b,\,\bar{K}'_b,\,f\longrightarrow l K_b,\,l\bar{K}'_b,\,lf$.)
On the other hand, the $\mubar$-scheme does not respect this scaling. In particular, the conditions for bounce occurrence given in \eqnref{eqn:qm bounce for pc} and \eqnref{eqn:qm bounce for pb} depend on $\mathcal{I}$ while those in \eqnref{eqn:qm' bounce for pc} and \eqnref{eqn:qm' bounce for pb} do not.

This implies that in the $\mubar$-scheme the choice of $\mathcal{I}$ has a physical consequence, and in the language of the no-hair theorem, one extra parameter ($M'$ or $K_c$) is required to completely  characterize the extended Schwarzschild solution. In the $\mubar'$-scheme, by contrast, the choice of $\mathcal{I}$ is physically irrelevant, and the no-hair theorem holds the same.

Additionally, the classical dynamics also admits the symmetries given by
\ba\label{eqn:symmetry 2}
\tau&\longrightarrow&\eta \tau,\nn\\
\gamma&\longrightarrow&\xi \gamma,\nn\\
p_b,\ p_c&\longrightarrow& \eta^2p_b,\ \eta^2p_c,\nn\\
b,\ c&\longrightarrow& \xi b,\ \xi c,\nn\\
K_c&\longrightarrow&\eta^2 K_c,\nn\\
M&\longrightarrow& \eta M.
\ea
The scaling symmetry regarding $\gamma\longrightarrow \xi \gamma$ is expected, since the Barbero-Immirzi parameter $\gamma$ has no effect on the classical dynamics. The scaling symmetry regarding $\tau\longrightarrow \eta \tau$ is also easy to understand, since there is no temporal scale introduced in the Hamiltonian.\footnote{For the Bianchi I cosmology studied in \cite{Chiou:2007mg}, a different scaling $p_I \longrightarrow p_I$ with $c_I\longrightarrow \eta^{-1}c_I$ is chosen to respect the symmetry regarding $\tau\longrightarrow \eta\tau$. This alternative scaling does not work in the case of Kantowski-Sachs spacetime, since it violates the Hamiltonian constraint \eqnref{eqn:cl rescaled Hamiltonian}. That is to say, the presence of the spatial curvature [i.e., the $\gamma^2p_b^2$ term in the bracket in \eqnref{eqn:cl rescaled Hamiltonian}] ties the temporal scale with the spatial scale; as a result, only the scaling $p_b,\,p_c \longrightarrow \eta^2p_b,\,\eta^2p_c$ (with gives the spatial direction the same scaling as in the temporal direction) with $b,\,c\longrightarrow b,\,c$ preserves the symmetry.} However, very surprisingly, the scaling symmetry involving $\tau\longrightarrow \eta \tau$ is violated for both the $\mubar$-scheme and the $\mubar'$-scheme phenomenological dynamics. Curiously, this symmetry is restored if $\tau\longrightarrow \eta \tau$ is accompanied by $\gamma \longrightarrow\xi \gamma$ and one extra scaling is also imposed at the same time:
\be\label{eqn:symmetry 2'}
\Delta\longrightarrow \xi^{-2}\eta^2\Delta.
\ee
This intriguing observation seems to suggest, albeit speculatively, that in the context of quantum gravity the fundamental scale (area gap) in spatial geometry gives rise to a temporal scale via the nonlocality of quantum gravity (i.e., using holonomies) and the Barbero-Immirzi parameter $\gamma$ somehow plays the role bridging the scalings in time and space. [This reminds us that, in LQG, the precise value of the area gap $\Delta$ is proportional to $\gamma$, and $\gamma$ is also the parameter which relates the \emph{intrinsic} geometry (encoded by spin connection $\Gamma^i_a$) with the \emph{extrinsic} curvature (${K_a}^i$) via ${A_a}^i=\Gamma^i_a-\gamma{K_a}^i$.]
Moreover, taking \eqnref{eqn:symmetry 2} and \eqnref{eqn:symmetry 2'} into \eqnref{eqn:qm' precise effective mass}, we also have
\be
\mathfrak{M}\longrightarrow\eta\mathfrak{M}.
\ee


The above observations for scaling symmetry draw close parallels to those in \cite{Chiou:2007mg} and \cite{Chiou:2008eg} for the phenomenological dynamics of LQC in the Bianchi I and Kantowski-Sachs models. Because of the absence of matter content, however, some implications thereof are missing here; particularly, the occurrence of bounces is no longer indicated by the (directional) matter energy density. Nevertheless, if we define ``energy density'' $\rho$ and ``directional densities'' $\rho_b$, $\rho_c$ as
\be\label{eqn:energy density}
\rho:=\frac{K_c^2}{8\pi Gp_b^2p_c},
\quad
\rho_b:=\frac{K_c^2}{8\pi Gp_b^3},
\quad
\rho_c:=\frac{K_c^2}{8\pi Gp_c^3},
\ee
then according to \eqnref{eqn:qm bounce for pc}, \eqnref{eqn:qm bounce for pb}, \eqnref{eqn:qm' bounce for pc} and \eqnref{eqn:qm' bounce for pb}, the bounces can still be said to take place whenever ``energy density'' or ``directional density'' approaches the Planckian density $\rho_{\rm Pl}:=(8\pi G\gamma^2\Delta)^{-1}$ (up to a numerical factor). This not only paraphrases the condition of bounce occurrence in a universal form as (generalized) energy density being the indicator for the bounce but also suggests that we should put the anisotropic shear on the equal footing as matter content and take into account the energy density arising from it.\footnote{As remarked in Section II.B of \cite{Chiou:2008eg}, the dynamics with Kantowski-Sachs symmetry closely resembles that in the Bianchi I model, implying that $K_c$ and $K_b$ characterize anisotropic shear and the Hamiltonian constraint can be understood as the relation which relates anisotropy with spatial curvature (and matter energy if any). Moreover, it has been shown in Appendix B of \cite{Chiou:2007sp} that the anisotropic shear behaves as a kind of anisotropic matter: the quantities defined in \eqnref{eqn:energy density} can be considered as the ``energy density of the classical anisotropic shear'' (portioned to the specific direction).} From this perspective, the ideas of relational interpretation of quantum mechanics remarked in \cite{Chiou:2007mg,Chiou:2008eg} can be carried over even without matter content.

Unfortunately, all the scaling symmetries break down in the detailed construction for the quantum geometry of the Schwarzschild interior. The fundamental quantum theory only respects the scaling symmetries at the leading order. This is due to the fact that the quantum evolution in the fundamental theory is governed by a difference equation, in which the step size of difference introduces an additional scale in the deep Planck regime \cite{Ashtekar:2005qt}. In fact, already in the level of phenomenological dynamics, the scaling symmetries are violated if we further take into account the loop quantum corrections on the cotriad component $\omega_c$. For the fundamental quantum theory, if we take the aforementioned symmetries seriously, we should revise the detailed construction to have the step size in the difference equation scale accordingly such that the symmetries are respected.

\section{Summary and discussion}\label{sec:summary}

\begin{figure}
\begin{picture}(500,230)(0,0)

\put(-63,-507)
{
\scalebox{1}{\includegraphics{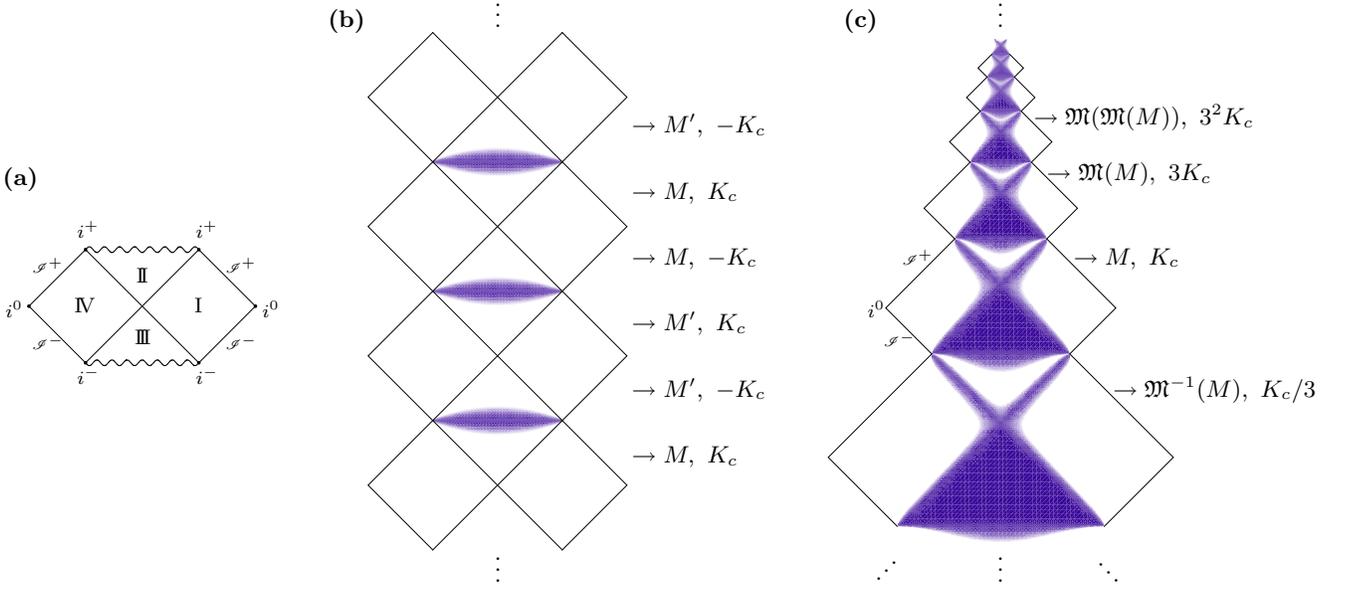}}
}

\end{picture}
\caption{\textbf{(Conjectured) Penrose diagrams.} \textbf{(a)} (Maximally extended) classical Schwarzschild spacetime. Region I\!I is the black hole; region I\!I\!I is the white hole; region I is the asymptotically flat region, external to the black hole; and region I\!V is the other asymptotically flat region. The wiggly lines are the black hole and the white hole singularities. \textbf{(b)} Schwarzschild spacetime in the $\mubar$-scheme. Both the classical black and white hole singularities are resolved by the quantum bounces (shaded areas), which bridge black holes with white holes. \textbf{(c)} Schwarzschild spacetime in the $\mubar'$-scheme. The classical black hole singularity is resolved and the event horizon is diffused by the quantum bounce. As a result, jumping over the quantum bounce (shaded area), the black hole gives birth to a baby black hole with increased $K_c$ and drastically decreased mass. This lineage continues until eventually $p_b$ grows exponentially and $p_c$ descends into a constant in the deep Planck regime as the spacetime becomes highly quantum mechanical. [The shaded areas indicate the regions where the quantum effects are significant (the darker the shade, the stronger the quantum effects). The patches for regions I and I\!V drawn in \textbf{(b)} and \textbf{(c)} are only conjectural.]}\label{fig:Penrose diagrams}
\end{figure}

To summarize, we list the important facts for the classical dynamics, $\mubar$-scheme and $\mubar'$-scheme phenomenological dynamics in \tabref{tab:summary}. The conjectured Penrose diagrams are depicted in \figref{fig:Penrose diagrams}. In the following, the main results are restated and their implications are discussed.


\begin{table}

\begin{tabular}{|c|c|c|}
\hline\hline
\rule[-2mm]{0mm}{6mm}
\textbf{Classical dynamics} & \textbf{Phenomenology in $\mubar$-scheme} & \textbf{Phenomenology in $\mubar'$-scheme}\\
\hline\hline

\begin{tabular}{c}
$2\pi p_b=2\pi L\sqrt{g_{xx}g_{\Omega\Omega}}={\bf A}_{x\phi}={\bf A}_{x\theta}$\\
$\pi p_c=\pi g_{\Omega\Omega}={\bf A}_{\theta\phi}$
\end{tabular}
&
\begin{tabular}{c}
$2\pi p_b=2\pi L\sqrt{g_{xx}g_{\Omega\Omega}}={\bf A}_{x\phi}={\bf A}_{x\theta}$\\
$\pi p_c=\pi g_{\Omega\Omega}={\bf A}_{\theta\phi}$
\end{tabular}
&
\begin{tabular}{c}
\rule{0mm}{3.5mm}
$2\pi p_b=2\pi L\sqrt{g_{xx}g_{\Omega\Omega}}={\bf A}_{x\phi}={\bf A}_{x\theta}$\\
$\pi p_c=\pi g_{\Omega\Omega}={\bf A}_{\theta\phi}$\vspace{1mm}
\end{tabular}\\

\hline
\begin{tabular}{l}
$b=\gamma\frac{d}{d\tau}\sqrt{g_{\Omega\Omega}}=\frac{\gamma}{2p_c^{1/2}}\frac{dp_c}{d\tau}$\\
$c=\gamma\frac{d}{d\tau}(L\sqrt{g_{xx}})$\\
\quad$=\frac{\gamma}{p_c^{1/2}}\frac{dp_b}{d\tau}-\frac{\gamma p_b}{2p_c^{3/2}}\frac{dp_c}{d\tau}$
\end{tabular}
&
\begin{tabular}{l}
$\frac{\sin(\mubar_bb)}{\mubar_b}=\frac{1}{\cos(\mubar_cc)}
\frac{\gamma}{2p_c^{1/2}}\frac{dp_c}{d\tau}$\\
$\frac{\sin(\mubar_cc)}{\mubar_c}=\frac{1}{\cos(\mubar_bb)}
\frac{\gamma}{p_c^{1/2}}\frac{dp_p}{d\tau}$\\
$\qquad\qquad-\frac{1}{\cos(\mubar_cc)}
\frac{\gamma p_b}{2p_c^{3/2}}\frac{dp_c}{d\tau}$
\end{tabular}
&
\begin{tabular}{l}
\rule[-3mm]{0mm}{8mm}
$\frac{\sin(\mubar'_bb)}{\mubar'_b}=\frac{1}{\cos(\mubar'_cc)}
\frac{\gamma}{2p_c^{1/2}}\frac{dp_c}{d\tau}$\rule{0cm}{0.45cm}\\
$\frac{\sin(\mubar'_cc)}{\mubar'_c}=\frac{1}{\cos(\mubar'_bb)}
\frac{\gamma}{p_c^{1/2}}\frac{dp_p}{d\tau}$\\
$\qquad\qquad-\frac{1}{\cos(\mubar'_cc)}
\frac{\gamma p_b}{2p_c^{3/2}}\frac{dp_c}{d\tau}
$\vspace{1mm}
\end{tabular}\\

\hline
\begin{tabular}{l}
$p_cc=\gamma K_c$\\
$p_bb=\gamma K_b(t')$
\end{tabular}
&
\begin{tabular}{l}
\rule{0mm}{4.5mm}
$p_c\frac{\sin(\mubar_cc)}{\mubar_cc}=\gamma K_c$\rule{0cm}{0.4cm}
\vspace{1mm}\\
$p_b\frac{\sin(\mubar_bb)}{\mubar_bb}=\gamma \bar{K}_b(t')$
\vspace{1mm}
\end{tabular}
&
\begin{tabular}{l}
$p_cc=\gamma\left[K_c+f(t'')\right]$
\vspace{1mm}\\
$p_bb=\gamma\left[\bar{K}'_b(t'')+f(t'')\right]$
\end{tabular}\\

\hline
$2K_bK_c+K_b^2+p_b^2=0$
&
$2\bar{K}_bK_c+\bar{K}_b^2+p_b^2=0$
&
\begin{tabular}{l}
\rule{0mm}{5mm}
$2\sin\left(\sqrt{\frac{\Delta\gamma^2}{p_b^2p_c}}(\bar{K}'_b+f)\right)
\sin\left(\sqrt{\frac{\Delta\gamma^2}{p_b^2p_c}}(K_c+f)\right)$
\vspace{1.5mm}\\
$\qquad\ +\sin^2\left(\sqrt{\frac{\Delta\gamma^2}{p_b^2p_c}}(\bar{K}'_b+f)\right)
+\frac{\Delta\gamma^2}{p_c}=0$
\vspace{1.5mm}
\end{tabular}\\

\hline
\begin{tabular}{l}
$\frac{1}{p_c}\frac{dp_b}{dt'}=\frac{{\bf V}}{4\pi p_c}\frac{dp_c}{d\tau}
=2K_b(t')$\\ \\ \\
$\frac{1}{p_b}\frac{dp_b}{dt'}=\frac{{\bf V}}{4\pi p_c}\frac{dp_b}{d\tau}
=K_b(t')+K_c$\\ \\ \\
$\frac{dK_b}{dt'}=\frac{{\bf V}}{4\pi}\frac{dK_b}{d\tau}=-p_b^2$
\end{tabular}
&
\begin{tabular}{l}
$\frac{1}{p_c}\frac{dp_c}{dt'}=2\cos(\mubar_cc)\bar{K}_b(t')$\\ \\ \\
$\frac{1}{p_b}\frac{dp_b}{dt'}=
\cos(\mubar_bb)\left[\bar{K}_b(t')+K_c\right]$\\ \\ \\
$\frac{d\bar{K}_b}{dt'}=
-\cos(\mubar_bb)\,p_b^2$
\end{tabular}
&
\begin{tabular}{l}
\rule{0mm}{5mm}
$\frac{1}{p_c}\frac{dp_c}{dt'}
=2\sqrt{\frac{p_b^2p_c}{\gamma^2\Delta}}
\cos\left(\sqrt{\frac{\gamma^2\Delta}{p_b^2p_c}}(K_c+f)\right)$\rule{0cm}{0.5cm}
\vspace{1.5mm}\\
$\qquad\qquad\times
\sin\left(\sqrt{\frac{\gamma^2\Delta}{p_b^2p_c}}(\bar{K}'_b+f)\right)$
\vspace{1.5mm}\\
$\frac{1}{p_b}\frac{dp_b}{dt'}
=\sqrt{\frac{p_b^2p_c}{\gamma^2\Delta}}
\cos\left(\sqrt{\frac{\gamma^2\Delta}{p_b^2p_c}}(\bar{K}'_b+f)\right)$
\vspace{1.5mm}\\
$\qquad\qquad\times\left[
\sin\left(\sqrt{\frac{\gamma^2\Delta}{p_b^2p_c}}(\bar{K}'_b+f)\right)\right.$
\vspace{1.5mm}\\
$\qquad\qquad\qquad\qquad+\left.\sin\left(\sqrt{\frac{\gamma^2\Delta}{p_b^2p_c}}(K_c+f)\right)
\right]$
\\
$\frac{d\bar{K}'_b}{dt'}=-p_b^2$
\vspace{1.5mm}
\end{tabular}\\

\hline
\begin{tabular}{l}
$p_c,\,p_b\rightarrow 0$\\
$\quad$toward classical singularity.\\ \\
$p_c\rightarrow 4G^2M^2,\quad p_b\rightarrow 0$\\
$\quad$toward event horizon.\\ \\ \\
\end{tabular}
&
\begin{tabular}{l}
$p_c$ bounces whenever\\
$\quad
p_c=\left(\gamma^2K_c^2\Delta\right)^{1/3}$;\\ \\
$p_b$ bounces whenever\\
$\quad
p_b\approx\left(4\gamma^2K_c^2\Delta\right)^{1/3}$.\\ \\
Epochs of bounces in $p_b$ and $p_c$\\
could be very separate.
\end{tabular}
&
\begin{tabular}{l}
$p_c$ bounces around the moment when\\
$\quad
\frac{\gamma^2\Delta}{p_b^2p_c}\approx
\frac{2(3-\sqrt{3}\,)}{(\bar{K}_b'-K_c)^2}
\approx
\left\{
\begin{array}{c}
\frac{2(3-\sqrt{3})}{9K_c^2},\\
\frac{2(3-\sqrt{3})}{K_c^2}.
\end{array}
\right.
$
\vspace{0.75mm}\\
$p_b$ bounces around the moment when\\
$\quad
\frac{\gamma^2\Delta}{p_b^2p_c}\approx
\frac{6}{(\bar{K}_b'-K_c)^2}
\approx
\left\{
\begin{array}{c}
\frac{2}{3K_c^2},\\
\frac{6}{K_c^2}.
\end{array}
\right.
$
\vspace{0.75mm}\\
$p_b$, $p_c$ bounce roughly around the\\
same moments.
\vspace{0.5mm}
\end{tabular}\\

\hline
\begin{tabular}{c}
No quantum bounce.\\ \\
$K_c$ fixed.\\
$M$ fixed.\\ \\ \\ \\
\end{tabular} &
\begin{tabular}{l}
Classical singularity is resolved by\\
the quantum bounce, which\\
bridges the classical black hole\\
with a classical white hole:\\
$\qquad\qquad\quad\, K_c\leftrightarrow -K_c$,\\
$\qquad\qquad\quad\ M\leftrightarrow M'$.
\\ \\
\end{tabular}
&
\begin{tabular}{l}
Classical singularity is resolved and\\
event horizon is diffused;\\
Quantum bounces conjoin classical cycles\\
of black holes:\\
$\quad\qquad\, \cdots\leftrightarrow
3^{-1}K_c \leftrightarrow
K_c \leftrightarrow 3K_c
\leftrightarrow\cdots$,\\
$\qquad \cdots\leftrightarrow
\mathfrak{M}^{-1}(M) \leftrightarrow
M \leftrightarrow  \mathfrak{M}(M)
\leftrightarrow\cdots$.
\\
Eventually, $p_c$ descends into deep Planck\\
regime while $p_b$ grows exponentially.
\vspace{0.5mm}
\end{tabular}\\

\hline
\begin{tabular}{c}
Symmetry of scaling:\\
$\tau\longrightarrow\eta \tau$\\
$\gamma\longrightarrow\xi \gamma$\\
$p_b,\ p_c\longrightarrow l\eta^2p_b,\ \eta^2p_c$\\
$b,\ c\longrightarrow \xi b,\ l\xi c$\\
$K_c\longrightarrow l\eta^{2}K_c$\\
$M\longrightarrow \eta M$\\
$\mbox{}$
\vspace{1mm}
\end{tabular}
&
\begin{tabular}{c}
Symmetry of scaling:\\
$\tau\longrightarrow\eta \tau$\\
$\gamma\longrightarrow\xi \gamma$\\
$p_b,\ p_c\longrightarrow \eta^2p_b,\ \eta^2p_c$\\
$b,\ c\longrightarrow \xi b,\ \xi c$\\
$K_c\longrightarrow \eta^{2}K_c$\\
$M\longrightarrow \eta M$\\
$\Delta\longrightarrow\xi^{-2}\eta^2\Delta$
\vspace{1mm}
\end{tabular}
&
\begin{tabular}{c}
Symmetry of scaling:\\
$\tau\longrightarrow\eta \tau$\\
$\gamma\longrightarrow\xi \gamma$\\
$p_b,\ p_c\longrightarrow l\eta^2p_b,\ \eta^2p_c$\\
$b,\ c\longrightarrow \xi b,\ l\xi c$\\
$(K_c+f)\longrightarrow l\eta^{2}(K_c+f)$\\
$M,\ \mathfrak{M}\longrightarrow \eta M,\ \eta\mathfrak{M}$\\
$\Delta\longrightarrow\xi^{-2}\eta^2\Delta$
\vspace{1mm}
\end{tabular}\\


\hline\hline
\end{tabular}
\caption{Summary of the classical dynamics, $\mubar$-scheme and $\mubar'$-scheme phenomenological dynamics.}\label{tab:summary}
\end{table}

In the $\mubar$-scheme phenomenological dynamics, the classical singularity is resolved and replaced by the quantum bounce, which bridges the black hole interior with the interior of a white hole. The black hole mass $M$ is different from the white hole mass $M'$ in general while the constant $K_c$ flips signs but its magnitude is unchanged.

On the other hand, in the $\mubar'$-scheme phenomenological dynamics, the classical black hole singularity is resolved and the event horizon is diffused by the quantum bounce. Jumping over the quantum bounce, the classical black hole with $K_c$ and $M$ gives birth to a baby black hole with $3K_c$ and the decreased mass $\mathfrak{M}(M)$ in the consecutive classical cycle. The baby black hole also brings forth its own baby and this scenario continues, giving the extended spacetime ``fractal'' structure, until eventually $p_b$ grows exponentially and $p_c$ asymptotes to a fixed value in the deep Planck regime, where the spacetime is essentially quantum mechanical and the semiclassical analysis could be questioned.

With regard to the finite sized interval $\mathcal{I}$ chosen to make sense of the Hamiltonian formalism, the phenomenological dynamics in the $\mubar$-scheme depends on the choice of $\mathcal{I}$. Particularly, given the black hole mass $M$, the exact value $M'$ of the conjoined white hole depends on $\mathcal{I}$. In the language of the no-hair theorem, two parameters $M$ and $M'$ (or alternatively, say $M$ and $K_c$) are required to completely characterize the (extended) Schwarzschild solution, although the information of $M'$ is hidden by the horizon and inaccessible (at least semiclassically) to the external observer. By contrast, the phenomenological dynamics in the $\mubar'$-scheme is completely independent of $\mathcal{I}$ as is the classical dynamics and the no-hair theorem remains unchanged.

In addition to the symmetry related to the choice of $\mathcal{I}$, both schemes admit additional symmetries of scaling, which are suggestive that the fundamental scale (area gap) in spatial geometry may give rise to a fundamental scale in temporal measurement. These symmetries, however, break down in the construction for the fundamental quantum theory.

While the $\mubar'$-scheme has the advantage that its phenomenological dynamics is independent of $\mathcal{I}$, the fundamental quantum theory of the Schwarzschild interior based on the $\mubar'$-scheme is difficult to construct. Both the $\mubar$- and $\mubar'$-schemes have desirable merits and it is still disputable which one (or yet another possibility) is more faithful to implement the underlying physics of loop quantum geometry. This issue is in the same status as that in the Bianchi I \cite{Chiou:2007mg} and Kantowsi-Sachs \cite{Chiou:2008eg} cosmological models. Hopefully, the detailed investigations in this paper on both schemes would help elucidate this issue. However, we should keep in mind that the validity of the phenomenological analysis remains to be justified. Some initial attempt has been made in \cite{Cartin:2006yv} to construct the semiclassical wave functions in the original $\mu_o$-scheme. It would be worthwhile to extend the previous work to the improved ($\mubar$- or $\mubar'$-) scheme and compare the results with those obtained here.

Meanwhile, it has been suggested \cite{Ashtekar:2005cj} and recently analyzed in detail for 2-dimensional black holes \cite{Ashtekar:2008jd} that quantum geometry effects may provide a possible mechanism for recovery of information that is classically lost in the process of Hawking evaporation, primarily because the black hole singularity is resolved and consequently the quantum spacetime is sufficiently larger than the classical counterpart. It would be very instructive to study the information paradox in the context of loop quantum geometry of the Schwarzschild black hole, as both resolution of the classical singularity and augmentation of spacetime have been observed at the level of phenomenological dynamics.

Additionally, in the $\mubar'$-scheme, the quantum effects not only resolve the singularity but also modify the event horizon. The fact that the event horizon is diffused may have an impact on the Hawking evaporation process. However, the homogeneous framework used on this paper only allows us to study the interior of the black hole and it is unclear how exactly the horizon is diffused and pieced together with the exterior (region I or I\!V in \figref{fig:Penrose diagrams}). In order to extend the results to cover the whole spacetime, the next step would be to apply the techniques described here to the inhomogeneous formulation of spherically symmetric loop quantum geometry such as developed in \cite{Campiglia:2007pr} and \cite{Bojowald:2004af}. This in turn could enable us to study the collapsing scenario of loop quantum black holes.

\acknowledgements{
This work was supported in part by the Eberly Research Funds of The Pennsylvania State University.}


\appendix

\section{Phenomenological dynamics in the $\mu_o$-scheme}\label{sec:muzero dynamics}
One of the virtues of the improved strategy ($\mubar$- or $\mubar'$-scheme) in both the isotropic and Bianchi I models of LQC is to fix the serious drawback in the original strategy ($\mu_o$-scheme), whereby the critical value of matter density $\rho_\phi$ (in the isotropic model) or of directional densities $\varrho_I$ (in the Bianchi I model) at which the bounce occurs can be made arbitrarily small by increasing the momentum $p_\phi$ of the matter field, thereby giving the wrong semiclassical behavior \cite{Ashtekar:2006wn,Chiou:2007dn,Chiou:2007mg}. In the case of the Schwarzschild interior, without the reference of matter content, it is not clear whether the $\mu_o$-scheme is problematic in regard to semiclassicality. For comparison, the phenomenological dynamics in the $\mu_o$-scheme is presented here.

In the phenomenological theory of the $\mu_o$-scheme, we take the
prescription to replace $c$ and $b$ with $\sin(\muzero_cc)/\muzero_c$ and $\sin(\muzero_bb)/\muzero_b$ by introducing the \emph{fixed} numbers $\muzero_c$ and $\muzero_b$ for discreteness. Analogous to \eqnref{eqn:qm Hamiltonian},
we have the effective (rescaled) Hamiltonian constraint:
\ba\label{eqn:qm0 Hamiltonian}
H'_{\mu_o}&=&
-\frac{1}{2G\gamma}
\left\{
2\frac{\sin(\muzero_bb)}{\muzero_b}\frac{\sin(\muzero_cc)}{\muzero_c}\,p_bp_c
+\left(\frac{\sin(\muzero_bb)}{\muzero_b}\right)^2 p_b^2
+\gamma^2p_b^2
\right\}.
\ea
To get an idea where the quantum corrections become appreciable, employing the classical solution given by \eqnref{eqn:cl sol of Kb}, \eqnref{eqn:cl sol of pc} and \eqnref{eqn:cl sol of pb}, we estimate the quantities $\muzero_bb$ and $\muzero_cc$:
\ba
\muzero_bb &=&\gamma\muzero_b \frac{K_b}{p_b}
\rightarrow
\left\{
\begin{array}{lcl}
\infty & & \quad\text{as }t'\rightarrow\infty,\\
\,0 & & \quad\text{as }t'\rightarrow-\infty,\\
\end{array}
\right.\\
\muzero_cc &=&\gamma\muzero_c \frac{K_c}{p_c}
\rightarrow
\left\{
\begin{array}{lcl}
\ \infty & & \quad\text{as }t'\rightarrow\infty,\\
\frac{\gamma\muzero_c K_c}{4G^2M^2} & & \quad\text{as }t'\rightarrow-\infty.\\
\end{array}
\right.
\ea
This suggests that the quantum corrections are significant near the classical singularity and negligible on the horizon provided that
\be\label{eqn:qm0 reg}
K_c\ll\frac{4G^2M^2}{\gamma\muzero_c},
\ee
which can always be satisfied if we choose $\mathcal{I}$ small enough for a given $M$.

The equations of motion are given by the Hamiltonian constraint $H'_{\mu_o}=0$ and Hamilton's equations:
\ba
\label{eqn:qm0 eom 3}
\frac{dc}{dt'}&=&\{c,H'_{\mu_o}\}=2G\gamma\,\frac{\partial\, H'_{\mu_o}}{\partial p_c}
=-2\gamma^{-1}p_b\frac{\sin(\muzero_bb)}{\muzero_b}
\frac{\sin(\muzero_cc)}{\muzero_c},\\
\label{eqn:qm0 eom 4}
\frac{dp_c}{dt'}&=&\{p_c,H'_{\mu_o}\}=-2G\gamma\,
\frac{\partial\, H'_{\mu_o}}{\partial c}
=2\gamma^{-1}p_bp_c\cos(\muzero_cc)
\frac{\sin(\muzero_bb)}{\muzero_b},\\
\label{eqn:qm0 eom 5}
\frac{db}{dt'}&=&\{b,H'_{\mu_o}\}=G\gamma\,\frac{\partial\, H'_{\mu_o}}{\partial p_b}
=-\gamma^{-1}p_c\frac{\sin(\muzero_bb)}{\muzero_b}\frac{\sin(\muzero_cc)}{\muzero_c}
-\gamma^{-1}p_b\left[\frac{\sin(\muzero_bb)}{\muzero_b}\right]^2
-\gamma p_b,\\
\label{eqn:qm0 eom 6}
\frac{dp_b}{dt'}&=&\{p_b,H'_{\mu_o}\}=-G\gamma\,
\frac{\partial\, H'_{\mu_o}}{\partial b}
=\gamma^{-1}p_b\cos(\muzero_bb)
\left[p_b\frac{\sin(\muzero_bb)}{\muzero_b}
+p_c\frac{\sin(\muzero_cc)}{\muzero_c}\right],
\ea
which follow
\be\label{eqn:qm0 dKc/dt'}
\frac{d}{dt'}\left[p_c\frac{\sin(\muzero_cc)}{\muzero_c}\right]=0
\qquad\Rightarrow\qquad
p_c\frac{\sin(\muzero_cc)}{\muzero_c}=\gamma K_c
\ee
and
\be\label{eqn:qm0 dKb/dt'}
p_c\frac{\sin(\muzero_bb)}{\muzero_b}=:\gamma K_b^o(t'),
\qquad
\frac{d K_b^o}{dt'}
=-\gamma^2p_b^2\cos(\muzero_bb).
\ee
These are exactly the same as \eqnref{eqn:qm dKc/dt'}--\eqnref{eqn:qm Kb} except that the discreteness variables $\mubar_c$ and $\mubar_b$ are now replaced by $\muzero_c$ and $\muzero_b$.

Therefore, exploiting the close resemblance between the $\mubar$-scheme and $\mu_o$-scheme, we can readily repeat the calculation we did in \secref{sec:mubar dynamics} and obtain the differential equations [cf. \eqnref{eqn:qm diff eq for Kb}--\eqnref{eqn:qm diff eq for pb}]:
\be\label{eqn:qm0 diff eq for Kb}
\frac{d K^o_b}{dt'}=\cos(\muzero_bb)\left(2K^o_bK^o_c+\bar{K}_b^2-K_\phi^2\right),
\ee
\ba
\label{eqn:qm0 diff eq for pc}
\frac{1}{p_c}\frac{dp_c}{dt'}&=&2\cos(\muzero_cc)K^o_b,\\
\label{eqn:qm0 diff eq for pb}
\frac{1}{p_b}\frac{dp_b}{dt'}&=&\cos(\muzero_bb)\left[K^o_b+K_c\right],
\ea
where
\ba\label{eqn:qm0 cosc}
\cos(\muzero_cc)&=&\pm\left[1-\sin^2\muzero_cc\right]^{1/2}
=\pm\left[1-\left(\frac{\gamma\muzero_cK_c}{p_c}\right)^{\!2}\right]^{1/2},\\
\label{eqn:qm0 cosb}
\cos(\muzero_bb)&=&\pm\left[1-\sin^2\muzero_bb\right]^{1/2}
=\pm\left[1-\left(\frac{\gamma\muzero_c K^o_b}{p_b}\right)^{\!2}\right]^{1/2},
\ea
which give the bouncing solution similar to that given in the $\mubar$-scheme phenomenological dynamics except that the exact conditions at which the bounces takes place are given differently by [cf. \eqnref{eqn:qm bounce for pc} and \eqnref{eqn:qm bounce for pb}]
\ba
\label{eqn:qm0 bounce for pc}
p_c&=&\gamma\muzero_c K_c\ll4G^2M^2,\\
\label{eqn:qm0 bounce for pb}
p_b&=&\gamma\muzero_b\abs{K^o_b}
\approx2\gamma\muzero_c K_c\ll8G^2M^2.
\ea

The phenomenological dynamics of the $\mu_o$-scheme closely resembles that of the $\mubar$-scheme. The classical singularity is resolved and replaced by the quantum bounce, which bridges a black hole interior with a white hole interior. The dynamics also depends on the choice of $\mathcal{I}$. The exact solution with $\muzero_b=\muzero_c=\delta$ can be found in \cite{Modesto:2006mx,Bohmer:2007wi}.

\end{document}